\documentclass[12pt,preprint]{aastex}
\pdfoutput=1
\usepackage{./emulateapj5}
\usepackage{natbib}
\usepackage{./onecolfloat}
\usepackage{./apjfonts}
\usepackage{./subfigure}  
\usepackage{color}

\newcommand{\hinv}{$h^{-1}$} 
\newcommand{\minv}{$h^{-1} M_{\odot}$} 
\newcommand{\nrsph}{${\rm NR_{SPH}}$}
\newcommand{\csfsph}{${\rm CSF_{SPH}}$}
\newcommand{\agnsph}{${\rm AGN_{SPH}}$}
\newcommand{\thcsph}{${\rm TH.C_{SPH}}$}
\newcommand{\nramr}{${\rm NR_{AMR}}$}
\newcommand{\csfamr}{${\rm CSF_{AMR}}$}

\newcommand{\chandra}{\emph {Chandra}}

\newcommand{\tmw}{$T_{\rm{MW}}$} 
\newcommand{\tsl}{$T_{\rm{SL}}$}

\newcommand{\bdm}{\begin{displaymath}} 
\newcommand{\edm}{\end{displaymath}}
\newcommand{\beq}{\begin{equation}} 
\newcommand{\eeq}{\end{equation}} 
\newcommand{\beqnarr}{\begin{eqnarray}}
\newcommand{\eeqnarr}{\end{eqnarray}}
\newcommand{\bit}{\begin{itemize}} 
\newcommand{\eit}{\end{itemize}} 
\newcommand{\ben}{\begin{enumerate}} 
\newcommand{\een}{\end{enumerate}}
\newcommand{\bfi}{\begin{figure}[htb]} 
\newcommand{\bpfi}{\begin{figure}[p]}
\newcommand{\barr}{\begin{array}}
\newcommand{\earr}{\end{array}}
\newcommand{\bec}{\begin{center}}
\newcommand{\eec}{\end{center}}
\newcommand{\bs}{\begin{sideways}}
\newcommand{\es}{\end{sideways}}

\shorttitle{Temperature distribution in  SPH and AMR}
\shortauthors{Rasia et al.}
\begin{document}
\twocolumn[%
\title{Temperature Structure of the Intra-Cluster Medium from SPH and AMR simulations}
\author{
Elena~Rasia\altaffilmark{1},
Erwin T. Lau\altaffilmark{2,3},
Stefano Borgani\altaffilmark{4,5,6},
Daisuke Nagai\altaffilmark{2,3,7},
Klaus Dolag\altaffilmark{8,9},
Camille Avestruz\altaffilmark{2,3},
Gian Luigi Granato\altaffilmark{5},
Pasquale Mazzotta\altaffilmark{10},
Giuseppe Murante\altaffilmark{5},
Kaylea Nelson\altaffilmark{7}, and
Cinthia Ragone-Figueroa\altaffilmark{5,11}
}
\affil{$^1$ Department of Physics, University of Michigan, 450 Church St., Ann Arbor, MI  48109, USA, rasia@umich.edu}
\affil{$^2$ Department of Physics, Yale University, New Haven, CT 06520, USA}
\affil{$^3$ Yale Center for Astronomy and Astrophysics, Yale University, New Haven, CT 06520, USA}
\affil{$^4$ Dipartimento di Fisica dell' Universit\`a di Trieste, Sezione di Astronomia, via Tiepolo 11, I-34131 Trieste, Italy}
\affil{$^5$ INAF, Osservatorio Astronomico di Trieste, via Tiepolo 11, I-34131, Trieste, Italy}
\affil{$^6$ INFN, Instituto Nazionale di Fisica Nucleare, Trieste, Italy}
\affil{$^7$ Department of Astronomy, Yale University, New Haven, CT 06520, USA}
\affil{$^8$ University Observatory Munich, Scheiner-Str. 1, D-81679 Munich, Germany}
\affil{$^9$ Max Planck Institut f\"ur Astrophysik, Karl-Schwarzschild-Str. 1, D-85748 Garching, Germany}
\affil{$^{10}$ Dipartimento di Fisica, Universit\`a di Roma Tor Vergata, via della Ricerca Scientifica, I-00133, Roma, Italy}
\affil{$^{11}$ Instituto de Astronom\'ia Te\'orica y Experimental, Consejo Nacional de Investigaciones Cient\'ificas y T\'ecnicas de la Rep\'ublica Argentina, Observatorio Astron\'omico, Universidad Nacional de C\'ordoba, Laprida 854, X5000BGR, C\'ordoba, Argentina}

\begin{abstract}
  Analyses of cosmological hydrodynamic simulations of galaxy clusters
  suggest that X-ray masses can be underestimated by 10\%--30\%. The
  largest bias originates by both violation of hydrostatic equilibrium (HE)
  and an additional temperature bias caused by inhomogeneities in the
  X-ray emitting intra-cluster medium (ICM). To elucidate on this
  large dispersion among theoretical predictions,
   we evaluate the degree of temperature structures in
   cluster sets simulated either with smoothed-particle-hydrodynamics (SPH) and adaptive-mesh-refinement
  (AMR) codes.
  We find that the SPH simulations produce larger temperature
  variations connected to the persistence of both substructures and their 
  stripped cold gas.
  This difference is more evident in nonradiative
  simulations, while it is reduced in the presence of radiative
  cooling. We also find that the temperature variation in radiative
  cluster simulations is generally in agreement with the observed one
  in the central regions of clusters.  
  Around $R_{500}$ the temperature inhomogeneities of the SPH simulations 
  can generate twice the typical HE mass bias of the AMR sample.
  We emphasize that a detailed understanding of the physical processes
  responsible for the complex thermal structure in ICM requires improved
  resolution and high sensitivity observations in order to extend the
  analysis to higher temperature systems and larger cluster-centric
  radii.
  
\end{abstract}
]

\section{Introduction}\label{sec:intro} 

A number of independent analyses on cosmological hydrodynamic
simulations of galaxy clusters consistently show that hydrostatic
equilibrium (HE) masses underestimate true masses by 10\%--30\%, the
exact value depending on the physics of the intracluster medium
(ICM), the hydrodynamic scheme, the radius within which the mass is
measured, and the dynamical state of the clusters
(\citealt{rasia.etal.04, piffaretti&valdarnini, jeltema.etal.08,
  ameglio.etal.09, lau.etal.09, nelson.etal.12,sembolini.etal.13, ettori.etal.13})

\citet[][hereafter R12\footnote{All acronyms referring to published
  papers are summarized here for convenience: R12 for
  \cite{rasia.etal.12}; N07 for \cite{nagai.etal.07b}; M10 for
  \cite{meneghetti.etal.10}; F13 for \cite{frank.etal.13}; and N14 for
  \cite{nelson.etal.14}}]{rasia.etal.12}, using synthetic \chandra\
observations of a set of massive clusters simulated with the
smoothed-particle hydrodynamics (SPH) {\tt GADGET} code, found ICM
temperature inhomogeneities to be responsible for 10\%--15\%
mass bias, which adds to a comparable bias associated
with the violation of HE \cite[see
also][]{rasia.etal.06}.  On the other hand, no significant
contribution to the mass bias associated with ICM thermal
inhomogeneities was found by \citet[][hereafter N07]{nagai.etal.07b,nagai.etal.07a}, who analyzed simulations
from the Eulerian adaptive-mesh refinement (AMR) code {\tt ART}, or by
\citet[][hereafter M10]{meneghetti.etal.10}, who investigated SPH
simulations including thermal conduction.  Temperature perturbations of the N07
sample are, indeed, verified to provide a negligible contribution
(less than 5\% within $R_{500}$\footnote{$R_{\Delta}$ is the radius of
  a sphere of mass $R_{\Delta}$ with a density $\Delta$ times above
  the critical density. In this paper, we consider $\Delta =$ 2500,
  500, 200. }) to X-ray temperature bias \citep{khedekar.etal.13}.  At
the same time, the presence of thermal conduction tends to homogenize
the temperature of the medium especially in massive systems
\citep{dolag.etal.04}.

A theoretical clarification of the mismatch on the X-ray mass bias is
quite timely after the reported conflict between the constraints on
cosmological parameters derived from primary cosmic-microwave-background anisotropies measured
by {\it Planck} and cluster number counts \citep{planck.etal.13XX}. A
suggested solution for the reconciliation of the two sets of
parameters seeks a bias on the X-ray masses as large as 40\%
\citep[see also][]{vonderlinden.etal.14}. The observational evaluation
of such a bias is often done by comparing the masses derived from
X-ray with those estimated through gravitational lensing, believed to
illustrate the true masses. Simulations, however, indicate that even
the gravitational lensing technique could be biased \citep[][M10;
R12]{becker_kravtsov.11} because of the triaxiality of the
cluster potential well or the presence of substructures located
either within the cluster or along the line of sight. These
complications affect individual observations and, therefore, generate
a significant scatter around the true mass. As a matter of fact, as of today
no clear convergence has been reached on the observational ratio
between X-ray and gravitational lensing masses
\citep{zhang.etal.08,mahdavi.etal.08,zhang.etal.10,jee.etal.11,foex.etal.12,
  mahdavi.etal.13,vonderlinden.etal.14,israel.etal.14}.

Because of the difficulties of establishing the amplitude of the X-ray mass
bias from observations, the interpretation of the aforementioned
theoretical disagreement can be done only by systematically analyzing
the temperature inhomogeneities present in the three samples of N07,
M10, and R12.  In this work, we group the simulated clusters into two sets
 generically labeled the SPH set and the AMR set (see Section~2).
Despite this naming choice, we would like
to stress that our analysis does not aim to be a code-comparison
project for which other conditions (such as common initial conditions)
need to be met \citep[e.g.][]{frenk.etal.99, oshea.etal.05, valdarnini.etal.12, power.etal.13}.

After the characterization of the inhomogeneities in the samples we
will investigate how our simulated data relate to observations.  Both
SPH and AMR simulations have already been shown to reproduce the
observed temperature profiles, at least outside the cluster core
regions \citep[see reviews by][and references
therein]{borgani_kravtsov,reiprich.etal.13}.  However, no detailed
comparison has been carried out so far for the small-scale ICM
temperature structure  mostly because of a lack of observational
measurements.  Fluctuations in density and temperature have been
measured only in a few nearby, dynamically disturbed clusters
\citep{bourdin.etal.08,bautz.etal.09,zhang.etal.09,gu.etal.09,bourdin.etal.11,churazov.etal.12,bourdin.etal.13,rossetti.etal.13,schenck.etal.14}.
Just recently, \cite{frank.etal.13} (hereafter F13) measured the
temperature distribution in the central region (within $R_{2500}$) 
of 62 galaxy clusters identified in the HIghest X-ray FLUx Galaxy Cluster Sample
\cite[HIFLUGCS][]{reip_boh}.  F13 analyzed the X-ray Multi-Mirror (XMM-Newton)
observations by adopting the smoothed-particle interference technique
\citep{peterson.etal.07}.  For each cluster, they built the
emission-measured temperature distribution, calculated its median and dispersion.

Simulations are described in Section~2. The analysis is
divided into three parts. First, we measure the temperature
variation and interpret the results by comparing the
performances of the SPH and AMR codes (Section~3). Second, we
investigate how temperature fluctuations are connected with density
fluctuations (Section~4). Finally, we compare the radiative simulations with
the observational data of F13 (Section~5). We discuss the impact that
temperature inhomogeneities have on the hydrostatic mass estimates
in Section ~6 and outline our conclusions in Section~7.

\section{Simulations}\label{sec:sim} 

In this work, we analyze the original cluster samples simulated with the SPH
technique from which the subsamples of R12 and M10 were extracted. 
In addition, we study four different implementations
of the ICM physics.  At the same time, we add about 85 clusters taken from Nelson et
al. 2014 (hereafter N14) to the 16 objects of N07. The new set is carried out with 
the same AMR code of N07 with the implementation of nonradiative physics.

Both the SPH and AMR sets assume a flat $\Lambda$-cold-dark-matter model with small
differences in the choice of cosmological parameters. The small changes are not
expected to affect our results. The SPH, N07, N14 simulations, respectively,
adopt: $\Omega_{\rm M} = $ 0.24, 0.3, 0.27 for the matter density parameter;
$\Omega_{\rm bar} = $ 0.040, 0.043, 0.047 for the baryon density; $H_0$ = 
72, 70, 70 km s$^{-1}$ Mpc$^{-1}$ for the Hubble constant at redshift
zero; and $\sigma_8 =$ 0.8, 0.9, 0.82 for the normalization of the
power spectrum on a scale of 8 $h^{-1}$ Mpc.

\subsection{Smoothed-Particle Hydrodynamics Sets}

The largest SPH set includes halos identified within 29
Lagrangian regions selected from a low-resolution $N$-body simulation of
volume equal to 1 (\hinv Gpc)$^3$ and resimulated at high resolution
\citep{bonafede.etal.11}.  Twenty-four of these regions are centered on the most
massive clusters of the parent $N$-body simulation, while the remaining are centered on group-size halos.
The size of the regions is such that no low-resolution 
contaminant dark-matter (DM) particle is found within five virial radii from the central halo.
Within all of the regions, further halos are identified
leading to $\sim 160$ as the total number of objects with mass $M_{\rm
  vir} > 3 \times 10^{13}$ \hinv $M_{\odot}$.  We limit this study 
to 49 systems with mass $M_{500}$ greater than $0.9 \times 10^{14}$
\hinv$ M_{\odot}$. This threshold corresponds to a mass-weighted 
temperature $T_{\rm MW} (<R_{500}) \approx
2$ keV when using the mass--temperature relation 
derived by \cite{planelles.etal.13} and \cite{fabjan.etal.11}.

The resimulations are carried out with the TreePM-SPH {\tt GADGET-3} code
\citep{springel05} with three different flavors for the ICM physics:
\begin{enumerate}
\item \nrsph: nonradiative physics with an entropy-conserving
  prescription for the SPH \citep{springel&herquist02} and artificial
  viscosity with the viscosity delimiter described by
  \citet{balsara95} and \citet{steinmetz96}
\item \csfsph: including radiative cooling, star formation, and
  feedback in energy and metals from supernovae. The radiative
  cooling, introduced as in \citet{wiersma.etal.09}, accounts for
  cosmic microwave background, UV/X-ray background radiation from
  quasars and galaxies \citep{haardt&madau01}, and metal cooling
  typical of an optically thin gas in photoionization equilibrium
  \citep{ferland.etal.98}. The star formation and evolution is treated
  via multiphase particles \citep{springel&hernquist03} with a
  coexisting cold and hot phases. Stars are distributed assuming the
  initial mass function of \citet{chabrier03} and evolve following the
  recipes of \citet{padovani&matteucci93}.  The kinetic feedback
  \citep{springel&hernquist03} released from the explosion of
  supernovae was implemented assuming velocity of the winds equal to
  $v_w= 500$ s$^{-1}$ km. The simulated clusters analyzed by R12 are a
  subsample of this set.
\item \agnsph: similar to the previous physics but also adding the
  feedback from active galactic nuclei (AGNs) resulting from gas
  accretion onto super-massive black holes \citep[SMBH; see][]{ragone.etal.13}.  
  The numerical scheme is largely based on that originally proposed by
  Springel et al.  (2005). It follows the evolution of SMBH particles
  whose dynamics are controlled only by gravity and whose mass grows
  by accretion from the surrounding gas or mergers with other
  SMBHs.  The accretion produces radiative energy with an efficiency of 0.2,
  of which 20\% is thermally given to the gas particles in the vicinity of
  the SMBH. Ragone-Figueroa et al.\ (2013) find that the
  original method by \cite{springel.etal.05} needs some modifications concerning: (1) the way SMBHs act as sinks of gas,
  (2) the strategy to place the SMBHs at the center of the
  hosting galaxy, and (3) how the radiative energy produced by accretion is returned to the interstellar medium. 
  These changes are essential in order to adapt the numerical scheme to the moderate
  resolution of cosmological simulations as well as to produce a sensitive
  coupling with the multiphase model adopted to treat star formation.
\end{enumerate}

Masses of dark matter and gas particles are $m_{\rm dm}= 8.47 \times 10^8$
\minv\ and $m_{\rm gas} = 1.53 \times 10^8$ \minv, respectively. The
adopted Plummer-equivalent softening length for gravitational force
is fixed to $\epsilon=5$ $h^{-1}$ kpc in physical units below redshift
$z=2$, and it is set to the same value in comoving units at higher
redshift.  The minimum SPH smoothing length is $0.5 \times \epsilon$.

In order to assess the effect of thermal conduction in SPH simulations, we
further analyze the nine main halos from \cite{dolag.etal.09} from which
the sample of M10 was extracted. The choices of particle mass and softening
are similar to the previous sets.
 \begin{enumerate}
 \item [4] \thcsph: among the objects listed in Table~2 of
   \cite{dolag.etal.09}, we specifically consider those labeled with
   the letter a.  The simulation sets studied here are csf and
   csfc. The former is equivalent to the treatment of the \csfsph\
   set, and the latter includes the effect of thermal conduction
   \citep{jubelgas.etal.04,dolag.etal.04}, characterized by a
   conductivity fixed to one-third the Spitzer conductivity of a
   fully ionized unmagnetized plasma.
 \end{enumerate}

\subsection{Adaptive-mesh Refinement simulations}

The AMR set includes the clusters at $z=0$ studied in N07 and N14. We
refer the reader to both papers for the details of the simulations.
Here we summarize their key properties. Simulations were carried out with the
adaptive-refinement tree{\tt ART} code
\citep{kravtsov.etal.97,rudd.etal.08}, a Eulerian code that
uses adaptive refinement in space and time and nonadaptive refinement
in mass \citep{klypin.etal.01} to achieve the dynamic range necessary
to resolve the cores of halos formed in self-consistent cosmological
simulations.

The 16 clusters from N07 are simulated using a uniform $128^3$ grid and eight
levels of mesh refinement in boxes of $120$~\hinv~Mpc
and $80$~\hinv~Mpc as sides, corresponding to peak spatial
resolution of about $3.66$~\hinv~kpc and $2.44$~\hinv~kpc,
respectively.  The DM particle
mass inside the virial radius is $m_{\rm dm} =
9.1\times 10^{8}$~\minv\ and $m_{\rm dm} = 2.7\times 10^{8}$~\minv\ for
the two box sizes, respectively, and external regions are simulated
with lower mass resolution.
The N07 clusters are simulated with two gas physics recipes: (1)
nonradiative gas physics (\nramr) and (2) galaxy formation physics
with metallicity-dependent radiative cooling, star formation, thermal
feedback from supernovae type Ia and II, and UV heating due to
cosmological ionizing background (\csfamr).

The cluster sample of N14 is simulated in a cosmological box of
$500$~\hinv~Mpc on a side, with a uniform $512^3$ grid and eight levels
of mesh refinement, corresponding to a maximum comoving spatial
resolution of $3.8$~\hinv~kpc. We identify and select $85$ cluster-size
 halos with $M_{500} \geq 2\times10^{14}$ \minv. The
virial regions surrounding the selected clusters are resolved with a DM
particle mass of $m_{\rm dm} = 1.0\times 10^9$~\minv\, corresponding to an
effective particle number of $2048^{3}$ in the entire box, and the external
regions are simulated with lower mass resolution. The clusters are
simulated with nonradiative gas physics only and are used as a control sample. The larger statistics 
validate results from the N07 nonradiative set.


\subsection{Exclusion of Cold Gas Particles.} 

Radiative cooling converts part of the gas from the hot and diffuse
phase to a cold and dense phase, resulting in a runaway cooling that
increases the amount of cooled baryons unless the process is regulated by
energy feedback by stars and black holes.  The gas most affected by
this process is located in the central regions of the central galaxies
or merging subhalos. Most of the gas in the cooled phase has
sufficiently low temperature and hence does not emit in the X--ray
band. However, a small fraction of it, being dense and having a
temperature of order of a few $10^6$ K, might form bright clumps visible
in soft X-ray images \citep{roncarelli.etal.06,
  zhuravleva.etal.13,vazza.etal.13,roncarelli.etal.13}.

Careful X-ray analysis on {\it Chandra}-like mock images requires the
detection of these clumps through a wavelength algorithm
\cite[e.g.][]{vikh.etal.98} as done in N07 and R12 \citep[see
also][]{rasia.etal.06, ventimiglia.etal.12, vazza.etal.13} and their consequent masking. 
 To extend the observational approach to the direct study
of simulated systems, other techniques, based on the density 
or volume of the gas elements, have been proposed in the literature. In
\citet{roncarelli.etal.06, roncarelli.etal.13}, the densest particles
of each spherical shell of constant width ($\sim 0.5 \times R_{200}$)
are excluded once their cumulative volume reaches 5\% of the total
particle volume in the shell. \citet{zhuravleva.etal.13}
suggested cutting all cells with gas density $\rho$ that satisfies 
the condition $\log \rho > \log \{\rho\} + 3 \sigma_{10, \rho}$, where
$\{\rho\}$ is the median of the density in the radial shell and
$\sigma_{10, \rho}$ is analogous to the standard deviation of the
log-normal distribution.  R12 proposed a different approach
that also takes into account the information on the temperature of the
cooled gas and eliminates all gas elements with $T < 3 \times 10^6
\rho^{0.25}$ where the temperature, $T$, is expressed in keV and the
gas density, $\rho$, in g cm$^{-3}$.  The slope of this relation is
linked to the effective polytropic index of the gas, and the value of
the normalization weakly depends on the cluster temperature (see
Appendix A of R12) and does not vary for the samples considered in
this paper.
Removing gas particles (in SPH) and cells (in AMR) according to this
criterion amounts to excluding less than 0.1\% of the gas volume,
about 10 times less than the amount selected by the other methods.

In the rest of this paper, the analysis on all radiative simulations (\csfsph, \agnsph, TH.C$_{\rm SPH}$, and \csfamr) is
performed after the exclusion of the cold gas, as done in R12. We find that this
criterion in addition to more effectively removing the multiphase gas,
also preserves the presence of merging small-group-size
substructures. This requisite is essential to comparing our simulations
to the observational data of F13, who analyzed the whole region within
$R_{2500}$ without applying any masking either on substructures or
on the core.  No gas has, instead, been removed in the nonradiative
simulation (\nrsph\ and \nramr).


\section{Temperature structure in simulations}\label{sec:results} 

\subsection{Measurements of ICM Temperature}

From the values of mass ($m$), density, and temperature of each gas
element (particle or cell), we compute the gas mass-weighted
temperature, $T_{\rm MW}$, and the spectroscopic-like temperature
\citep{mazzotta.etal.04, vikh.etal.06b}, $T_{\rm SL}$, defined as
\begin{equation}
T=\frac{\Sigma W_i  T_i}{\Sigma W_i} \ \ \ \ {\rm with}\  \ \  \ \ W_{i,\rm MW}= m_i , \ \ {\rm or}  \ \ \ W_{i,\rm SL}= m_i \rho_i T_i^{-0.75}.
\end{equation}
In the above equation, the summation signs run over the gas elements
belonging to three regions: the innermost is the sphere with radius
$R_{2500}$ and the intermediate and outermost regions are spherical
shells delimited by [$R_{2500}$, $R_{500}$] and by [$R_{500}$,
$R_{200}$].  We label these regions {\it I}, {\it M}, and {\it O},
respectively.

\tmw\ has a well-defined physical meaning because it is directly related to
the total thermal energy of the ICM: $E_{\rm th} \sim m \times$ \tmw. As
such, it is the temperature that should be entered into the
HE mass estimate.  \tsl, on the other hand, is the
temperature directly accessible to X-ray spectroscopy, and it is more
sensitive to dense gas than \tmw.  In our analysis we include all
particles or cells with temperature above 0.5 keV
\citep{mazzotta.etal.04}.  For a thermally uniform medium these two
temperatures coincide. Any difference between
them, $\Delta_T=$ \tmw $-$ \tsl $>0$, could be interpreted as a
quantitative measure of inhomogeneities in the ICM thermal structure.
As matter of fact, the relation \tsl $<$ \tmw\ is verified whenever
cold and dense gas is within the considered region.
The opposite situation, \tmw $<$ \tsl, exists in the presence of a
negative temperature gradient aligned with a positive mass gradient,
as usually is the case in the outskirts of relaxed objects.

The dependence of $\Delta_T$ on $T_{\rm MW}$ for the SPH
and AMR codes is shown in Figure~1.
Tables~1 and ~3 report the best-fit linear relations to
\tmw-$\Delta_T$ along with the associated intrinsic scatter
$\sqrt{\chi^2/(N-1)}$, where $N$ is the number of clusters
analyzed. The best fits are computed by minimizing the $\chi^2$ error
statistic.

\subsection{Smoothed-Particle Hydrodynamics}

\begin{figure*}[ht!]
\centering
\includegraphics[width=0.4\textwidth]{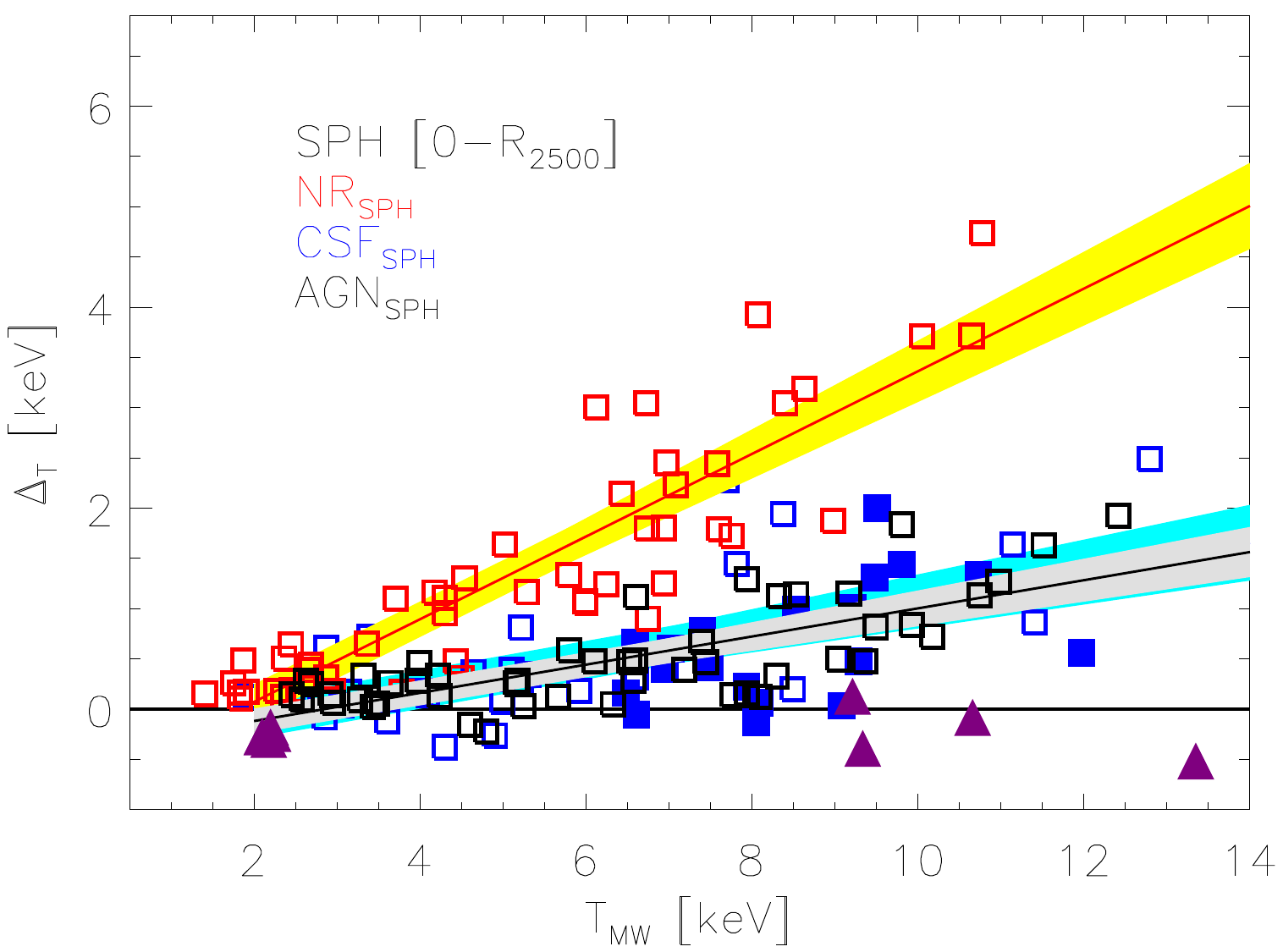}
\includegraphics[width=0.4\textwidth]{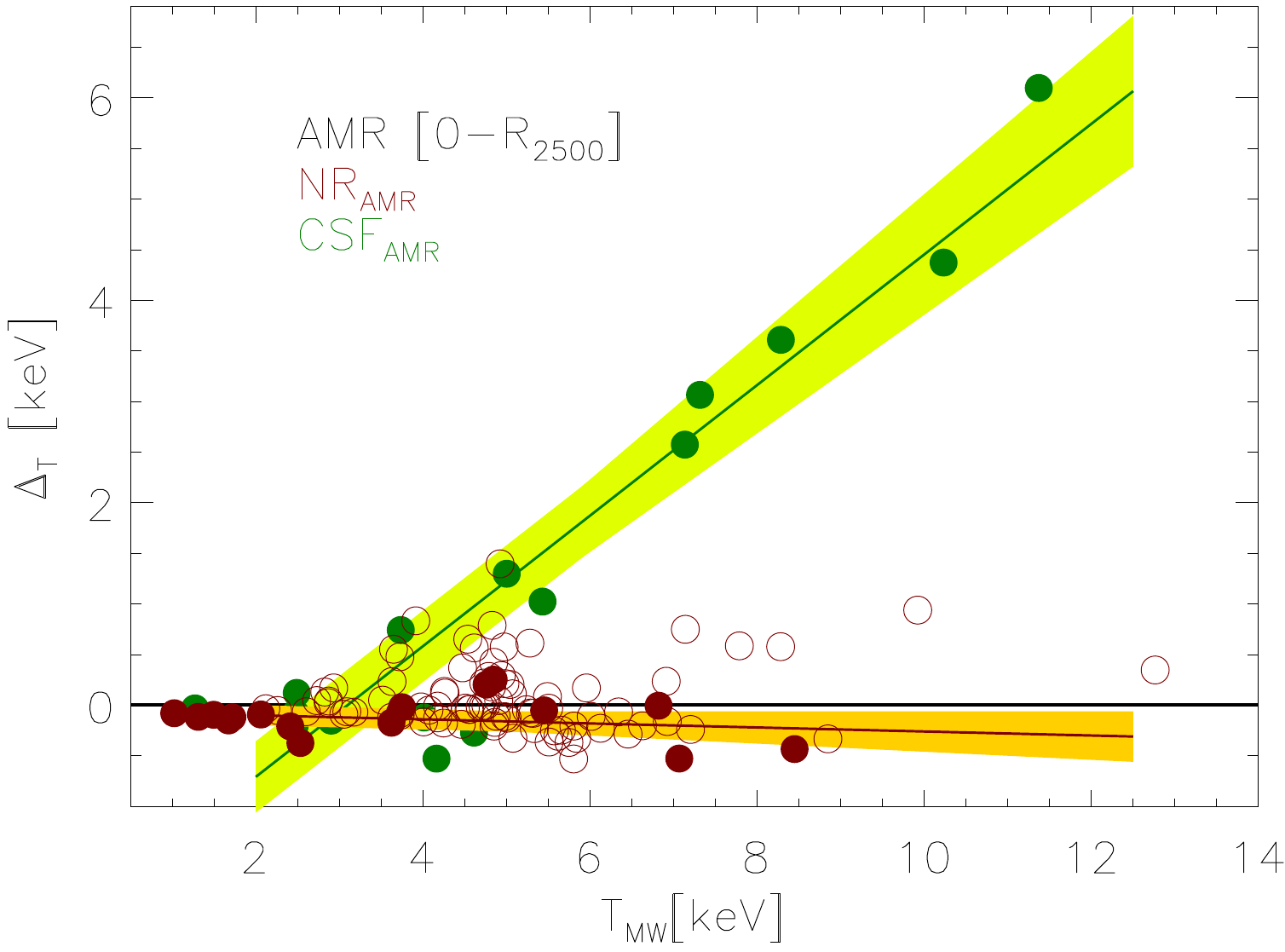}
\includegraphics[width=0.4\textwidth]{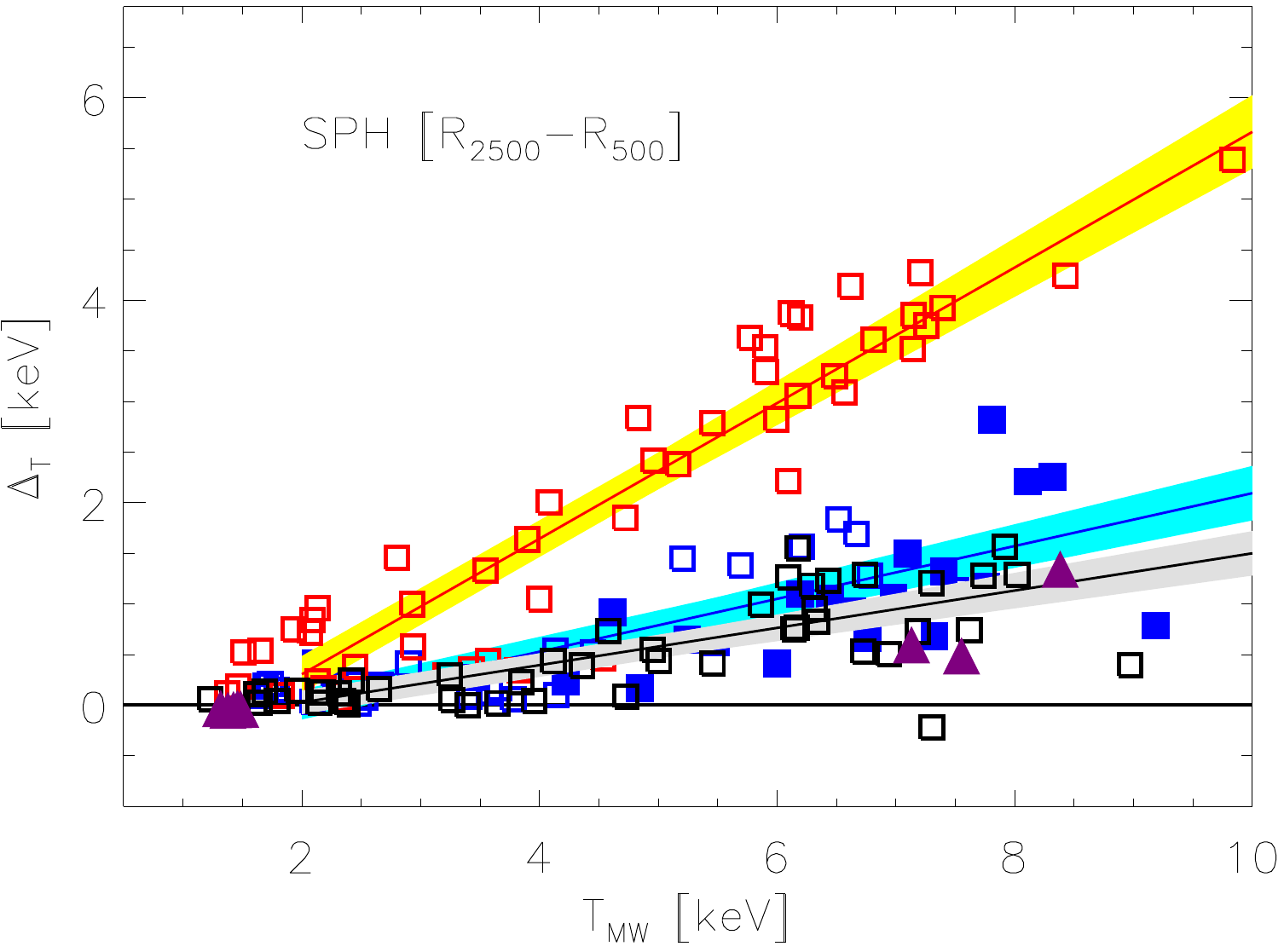}
\includegraphics[width=0.4\textwidth]{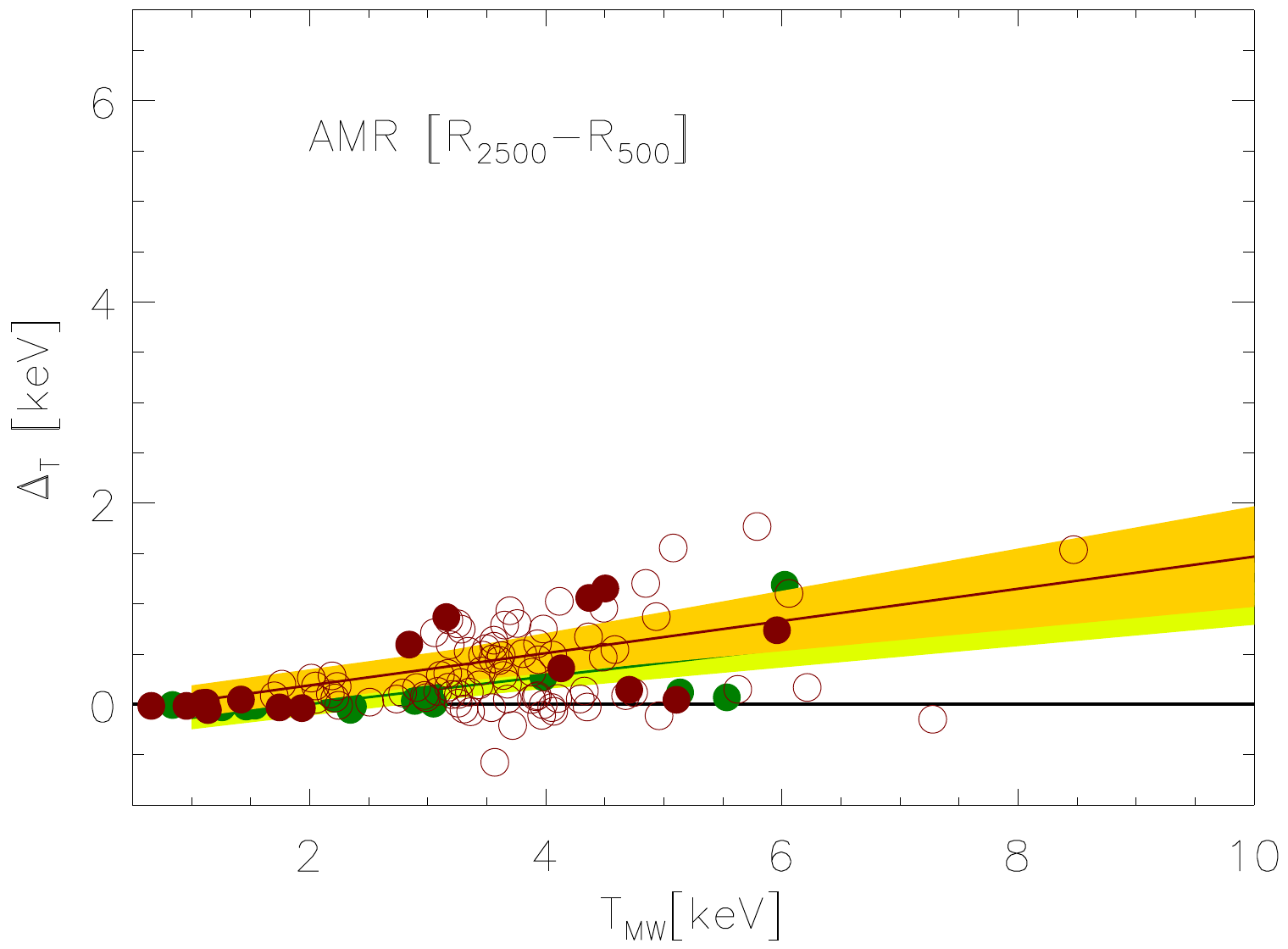}
\includegraphics[width=0.4\textwidth]{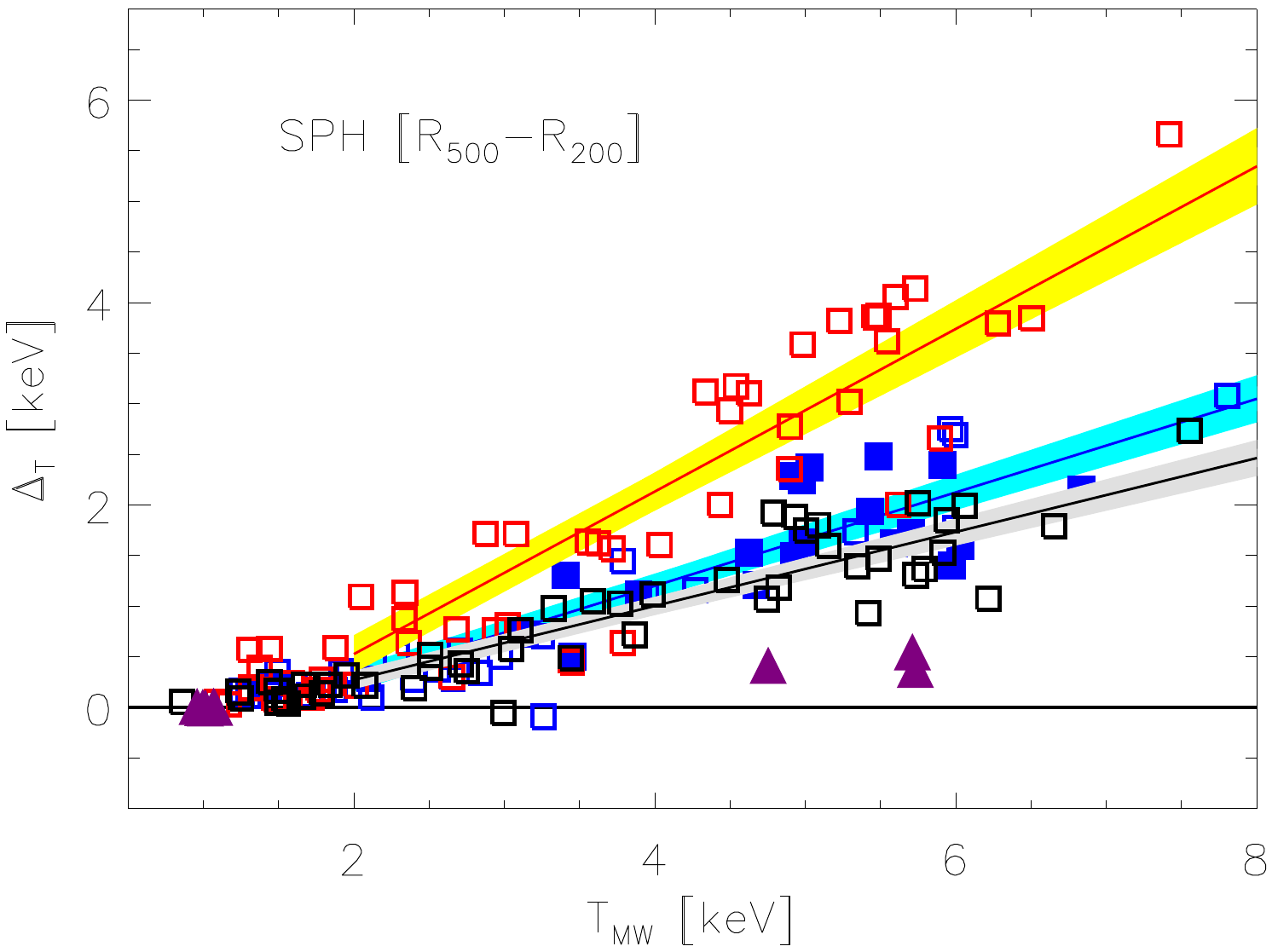}
\includegraphics[width=0.4\textwidth]{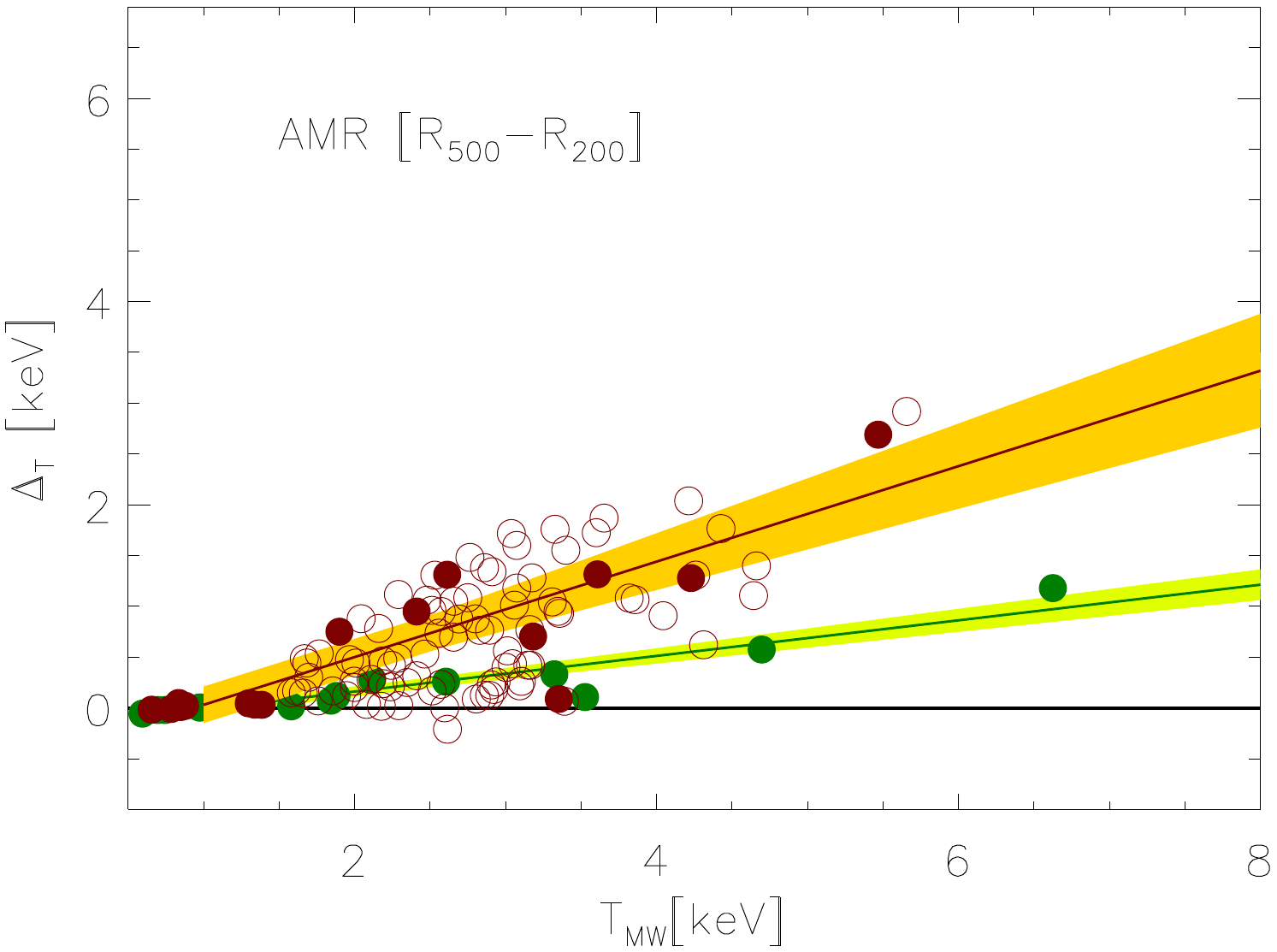}
\caption{Temperature variation, $\Delta_T=$ \tmw$-$\tsl, vs. \tmw\ for
  the SPH (left panels) and AMR (right panels) simulations measured in the {\it I} (top panel), {\it M} (center panel), and {\it O} regions (bottom panel).  Left panels: red, blue, and
  black squares correspond to \nrsph, \csfsph, and \agnsph,
  respectively, and purple triangles refer to ${\rm TH.C_{SPH}}$.  Filled squares indicate the subset analyzed by
  R12. Right panels: brown and
  green circles correspond to \nramr\ and \csfamr, respectively.
  Filled circles indicate the subset analyzed by N07. 
  The shaded regions represent the 1$\sigma$ scatter around the
  best-fitting relations. For clarity, we omit the best-fit relation of the thermal conduction simulations (see Table~1). }
\label{fig:dt_t}
\end{figure*}

The left panels of Figure~1 show the temperature variation, $\Delta_T$, measured in the SPH simulations.  As a general trend, the
nonradiative physics presents the largest degree of temperature variation at all
radii and along the entire temperature range. \csfsph\ clusters behave
similar to \agnsph\ in the innermost region but are typically above in the more external shells. 
${\rm
  TH.C_{SPH}}$ objects are characterized by a small variation in
temperature $\Delta_T<0.5$ keV. The differences between radiative and nonradiative simulations are
expected because the cooling process preferentially removes low-entropy gas
(especially that associated with central galaxies and merging
substructures) from the diffuse phase. This phenomenon, combined
with the heating provided by supernovae, decreases the temperature
contrast between  clumps and diffuse ICM.  In the
intermediate and outermost regions, the slope of the \tmw$ - \Delta_T$
relation for the CSF$_{\rm SPH}$ simulations decreases by 40--60\% with respect to
the NR$_{\rm SPH}$ case (Table~1).

The increase of the \tmw$ - \Delta_T$ slope moving
outward is expected because the outskirts of the most massive systems
are more severely affected by inhomogeneities generated by ongoing gas
accretions along filaments.  This picture is consistent
with the increase of the clumpy factor with radius
\citep{nagai&lau11,vazza.etal.13,khedekar.etal.13,
  roncarelli.etal.13}.  The accreting clumps are larger and survive
longer in nonradiative simulations \citep{dolag.etal.09}, affecting more
strongly the spectroscopic-like temperature
without significantly influencing the mass-weighted temperature.  
This statement is illustrated in the left panel of Figure~2 where we
show how the spectroscopic-like temperature and the mass-weighted
temperature change in radiative simulations with respect to nonradiative
simulations.  On the other hand, the imprint of the
particular feedback model (either by supernovae or by AGNs) has less impact on the calculation of both temperatures (right panel of
Figure~2). 
Thermal conduction minimally influences small clusters
but induces 50\% - 100\% variations in the $T_{\rm SL}$ measurements of the four most
massive systems. That said, the slope of the \tmw $- \Delta_T$ is still a factor of two-to-three lower than the
\csfsph\ case.

\begin{figure}[ht!]
\centering
\includegraphics[width=0.45\textwidth]{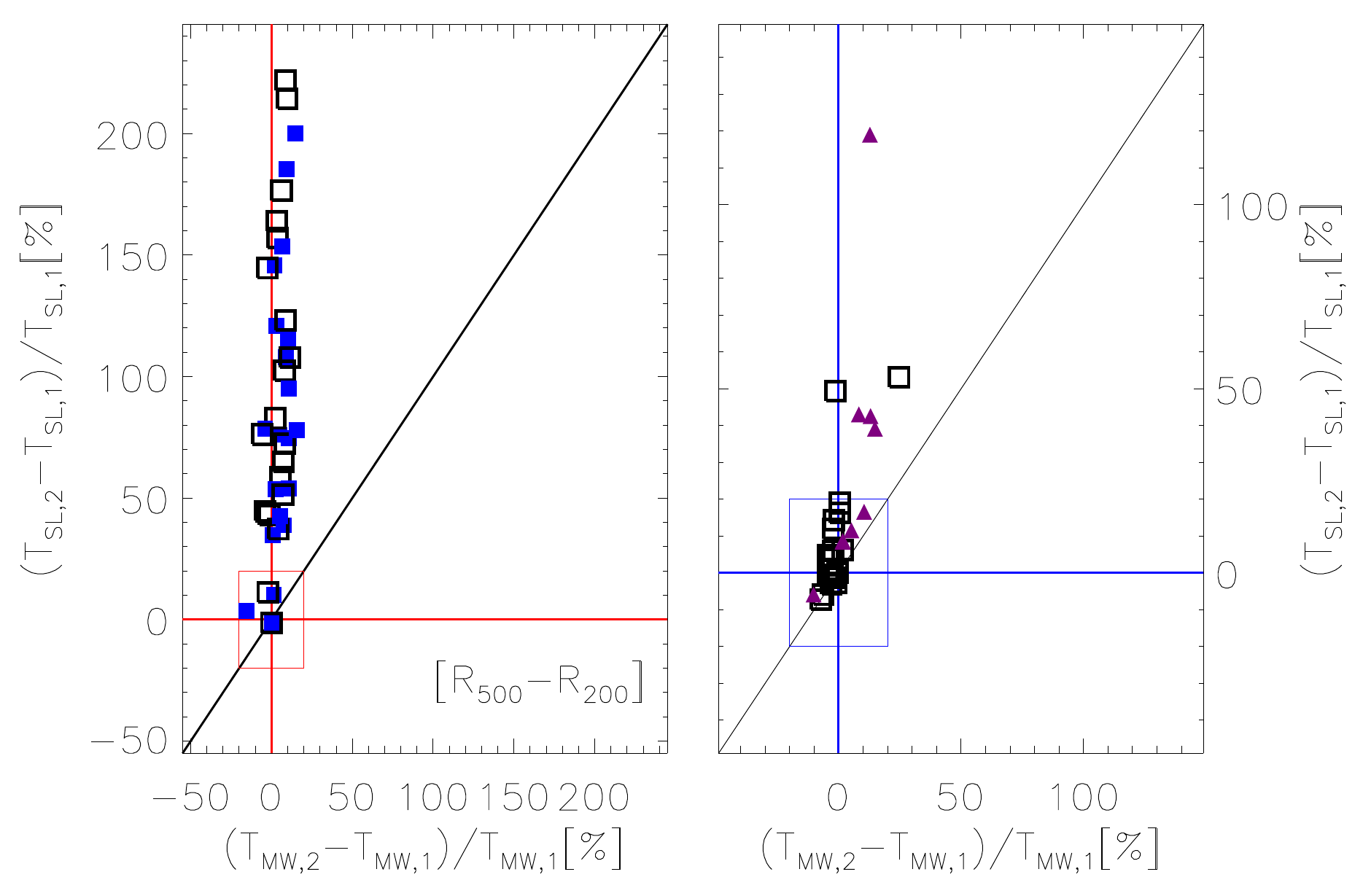}
\caption{{Difference between the two temperatures (mass-weighted and
    spectroscopic-like) as measured within the various SPH
    simulations in the {\it O} region. Left panel: the radiative cases (\csfsph\ in blue and \agnsph\ in
    black) of the R12 sample are compared to the \nrsph\ flavor. Right panel: \agnsph\ is
    related to \csfsph\ and ${\rm TH.C_{SPH}}$ is compared with the respective CSF$_{\rm SPH}$ physics. Symbols are as in Figure~1. The box in the center represents a 20\% variation in both panels.}}
\end{figure}

\begin{table}[ht!]
\caption{SPH Set: Best-fit Parameters, their 1 $\sigma$ Errors, and Scatter of the Linear Relation: $\Delta_T$=\tmw--\tsl$=A+B\times $(\tmw) }
\centering
\begin{tabular}{|lccc|} 
\hline
\hline
\multicolumn{4}{|c|}{all clusters}\\
 & A$\pm$ err(A) &B $\pm$ err(B) & scatter \\
\hline
\hline
\nrsph, I       & -0.75 $\pm$ 0.18 & 0.41$\pm$ 0.03  & 0.55 \\
\nrsph, M     &-1.03 $\pm$ 0.18 & 0.67 $\pm$ 0.04 & 0.54\\
\nrsph, O     &-1.08 $\pm$ 0.19 & 0.80 $\pm$ 0.05 & 0.56\\
\hline
\csfsph, I     &  -0.39 $\pm$ 0.19 &0.15 $\pm$ 0.023& 0.53 \\
\csfsph ,M   &  -0.52 $\pm$ 0.15 & 0.26 $\pm$ 0.03 & 0.43 \\
\csfsph, O   &  -0.65 $\pm$ 0.12 & 0.46 $\pm$ 0.03 & 0.35\\
\hline
\agnsph, I     & -0.40 $\pm$ 0.12 &0.14 $\pm$ 0.02 & 0.34 \\
\agnsph ,M   & -0.40 $\pm$ 0.12 & 0.19 $\pm$ 0.02 & 0.36 \\
\agnsph, O   & -0.50 $\pm$ 0.09 & 0.37 $\pm$ 0.02 & 0.28 \\
\hline
${\rm TH.C_{SPH}}$, I     & -0.22 $\pm$ 0.11 &  0.0 $\pm$  0.01&  0.18\\ 
${\rm TH.C_{SPH}}$ ,M   & -0.19 $\pm$ 0.13 &  0.12$\pm$ 0.02 &  0.22\\ 
${\rm TH.C_{SPH}}$, O   & -0.15 $\pm$ 0.05 &  0.13$\pm$ 0.01 & 0.10 \\ 
\hline
\multicolumn{4}{|c|}{only relaxed}\\
 & A$\pm$ err(A) &B $\pm$ err(B) & scatter \\
\hline
\nrsph, I       & -1.01 $\pm$ 0.29 & 0.38$\pm$ 0.05  & 0.40 \\
\nrsph, M     & -1.69 $\pm$ 0.47 & 0.76 $\pm$ 0.10 & 0.66\\
\nrsph, O     & -1.56 $\pm$ 0.41 & 0.90 $\pm$ 0.011 & 0.56\\
\hline
\csfsph, I     & -0.37 $\pm$ 0.16   &0.06 $\pm$ 0.02 & 0.21 \\
\csfsph ,M   &  -0.33 $\pm$ 0.27 & 0.17 $\pm$ 0.05 & 0.37 \\
\csfsph, O   &  -1.00 $\pm$ 0.18 & 0.56 $\pm$ 0.05 & 0.24\\
\hline
\agnsph, I     & -0.71 $\pm$ 0.24 &0.15 $\pm$ 0.04 & 0.30 \\
\agnsph ,M   & -0.17  $\pm$ 0.13 & 0.09 $\pm$ 0.02 & 0.17 \\
\agnsph, O   & -0.55 $\pm$ 0.20 & 0.37 $\pm$ 0.05 & 0.26 \\
\hline
\end{tabular}
{\footnotesize  Relaxed clusters are defined in Section 6.1}
\label{tab:relation}
\end{table}

The difference between the two temperatures, \tmw\ and \tsl, measured in
the \csfsph\ simulation is similar to those reported by
\citet{biffi.etal.14}, who analyzed about 180 massive clusters selected
from the ``Mare-Nostrum Multidark Simulations of Galaxy Clusters'' and
resimulated with the code {\tt GADGET}, including the treatment of
cooling, star formation, and feedback by supernovae. In that case, the
average difference between $T_{\rm MW}$ and $T_{\rm SL}$ is $\sim 15\%$ when
both measurements are carried out within a sphere of radius $R_{500}$. In our case, the
median of the ratios varies from 6\% in the {\it I} region to 14\% and
to 31\% in the $M$ and $O$ shells. The inclusion of AGN does not
change the value in the center but decreases the differences to 10\%
and 23\% in the other two regions. As expected, major variations
are present for the nonradiative simulations: the medians of the
ratios are about 20\% in the innermost sphere but exceed 70\%
elsewhere. This last value is significantly higher than the 20\%
predicted by the pioneering work of \cite{mathiesen_evrard}. A direct
comparison is, however, arduous because (1) the cosmological models adopted
there consider a higher spectrum normalization, $\sigma_8=1$,
significantly altering the evolution of large structures such as
galaxy clusters and (2) the simulated particle mass resolution is three orders of magnitude smaller than ours.

The profile of the temperature variation, $\Delta_T$, computed within
logarithmically spaced spherical shells is shown in Figure~3 for the
four sets of SPH simulations.  The impact of merging
substructures is evident in the NR simulation profiles that show
significant temperature variations throughout the clusters: over 30\%
of the systems have $\Delta_T>2$ keV in the region $r\gtrsim 0.2
R_{200}$.

\begin{figure}[ht!]
\centering
\includegraphics[width=0.45\textwidth]{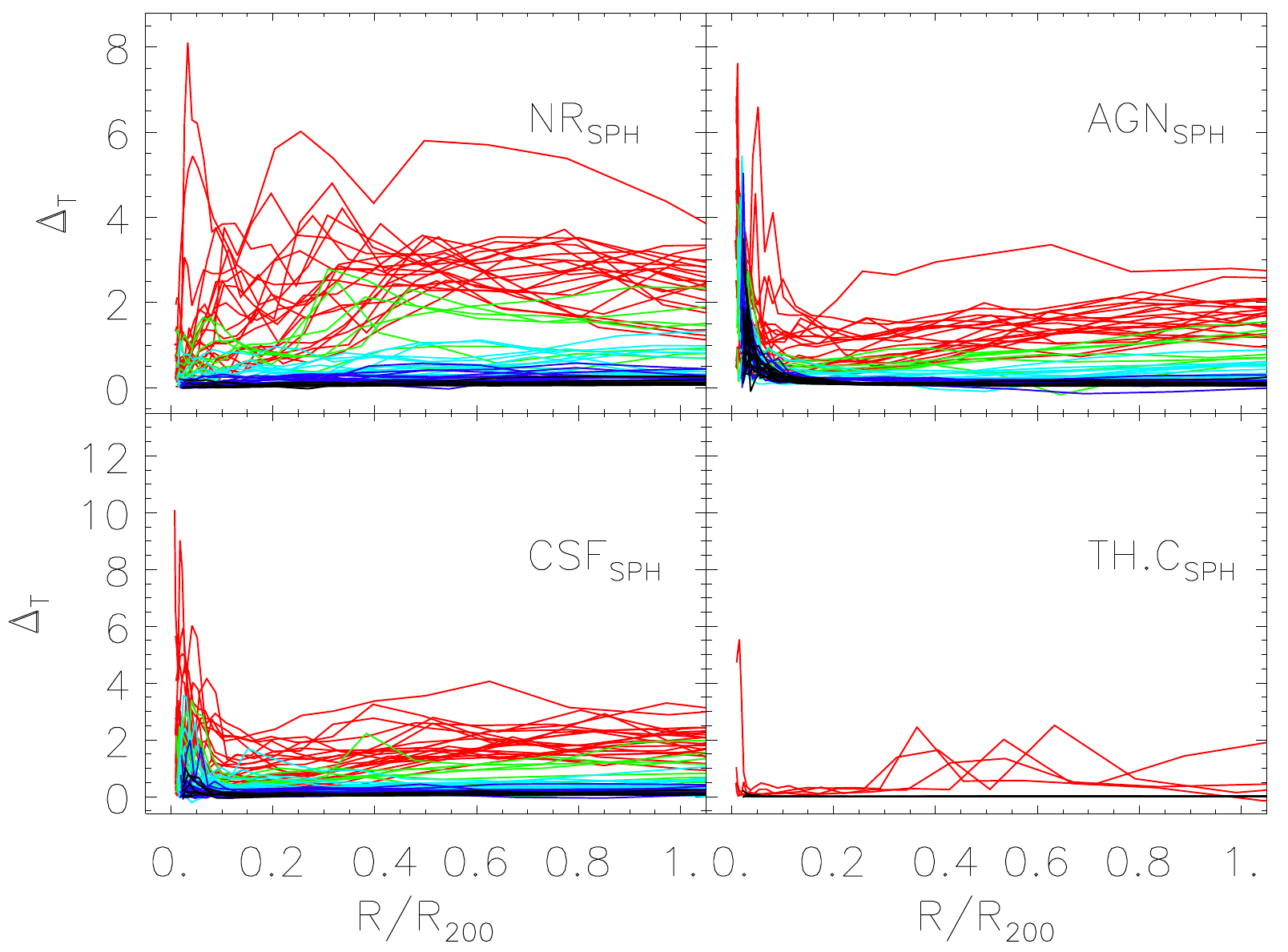}
\caption{Profiles of temperature variation, $\Delta_T=$ \tmw $-$ \tsl, for
  the different sets of simulated SPH clusters. Line color varies with
  cluster mass: 
  black lines for $M_{200}<10^{14}$ \minv;
  blue for $M_{200}$ in the range $[1 \div 2] \times 10^{14}$ \minv;
  cyan for $M_{200}$ in the range $[2 \div 5] \times 10^{14}$ \minv;
  green for $M_{200}$ in the range $[5 \div 10] \times 10^{14}$ \minv;
   and red for $M_{200}>10^{15}$ \minv.}
\end{figure}
This fraction is reduced in radiative simulations as a consequence
of the gas cooling mentioned above.  By comparing the simulations with and
without AGN feedback, we find that AGNs reduce the temperature variation by 35\%.
 
Thermal conduction, on the other hand, almost completely homogenizes
the ICM temperature structure. Even the profiles of the most massive
clusters are essentially flat and consistent with no
perturbations. The infall of substructures generates only localized
peaks but it does not cause long-lasting consequences for
the thermal structure of the diffuse medium. $\Delta_T$ is always below 0.5 keV at
$R_{200}$ with the exception of one object that is experiencing a merging at $z=0$. In
addition, temperature inhomogeneities are strongly reduced in the
central regions, including the core.  This behavior explains the
absence of a significant contribution of the temperature bias to the
measurement of X-ray mass bias as found by M10.

\subsection{Adaptive-mesh Refinement}

In the right column of Figure~1, we report the temperature variation as a function of the
mass-weighted temperature for the {\it I} and {\it O} regions of the
AMR clusters. Focusing first on the nonradiative results (red line)
allows us to evaluate the different predictions of the Eulerian code  with
respect to the Lagrangian one.  The \nramr\ slopes are always
shallower by 40\%-100\% (Table~2) than the \nrsph\ slopes (Table~1).
This could be explained by the larger amount of mixing in the AMR
code, which makes the stripping of low-temperature and loosely bound
cells more effective. This leads to dissolution of merging
substructures and reduction of gaseous inhomogeneties toward the inner
region of the clusters.  Within $R_{2500}$, indeed, \nramr\ simulations show
the equality between \tmw\ and \tsl\ ($B\sim 0$ in Table~2), 
suggesting that the ICM is thermally homogeneous.
On the other hand, the SPH simulations show larger temperature variations
because merging substructures are more persistent to disruption due to
the lack of thermal diffusion \citep[e.g.,][and references
therein]{frenk.etal.99,oshea.etal.05,power.etal.13}.  As in the
particle-based codes, the slope of the \tmw$-\Delta_T$ relation increases
in moving from the {\it I} region to the {\it O} (Table~2).

\begin{table}[ht!]
\caption{AMR Set: Best-fit Parameters, their 1 $\sigma$ Errors, and Scatter of
  the Linear Relation $\Delta_T$=\tmw-\tsl$=A+B\times $(\tmw) } \centering
\begin{tabular}{|lccc|} 
\hline
\hline
\multicolumn{4}{|c|}{all clusters}\\
 & A$\pm$ err(A) &B $\pm$ err(B) & scatter \\
\hline
\hline
\nramr, I     & -0.06 $\pm$ 0.10  & -0.02 $\pm$ 0.02 & 0.20\\
\nramr, M   & -0.13 $\pm$ 0.16 &  0.16 $\pm$ 0.05  & 0.33 \\
\nramr, O   & -0.44 $\pm$ 0.18  & 0.47 $\pm$ 0.07  & 0.38\\
\hline
\csfamr, I   & -2.0 $\pm$ 0.35 & 0.65 $\pm$ 0.06 & 0.65 \\
\csfamr, M & -0.27  $\pm$ 0.11 & 0.14 $\pm$ 0.03 & 0.23\\
\csfamr, O & -0.19 $\pm$ 0.05  & 0.18 $\pm$ 0.02 & 0.12\\
\hline
\hline
\multicolumn{4}{|c|}{only relaxed}\\
 & A$\pm$ err(A) &B $\pm$ err(B) & scatter \\
\hline
\hline
\nramr, I     & -0.05 $\pm$ 0.06  & -0.07 $\pm$ 0.02 & 0.09\\
\nramr, M   & -0.03 $\pm$ 0.17 &  0.05 $\pm$ 0.07  & 0.24 \\
\nramr, O   & -0.01 $\pm$ 0.21  & 0.09 $\pm$ 0.13  & 0.29\\
\hline
\csfamr, I   & -1.95 $\pm$ 0.36 & 0.60 $\pm$ 0.07 & 0.43 \\
\csfamr, M & -0.05  $\pm$ 0.04 & 0.02 $\pm$ 0.01 & 0.04\\
\csfamr, O & -0.03 $\pm$ 0.03  & 0.05 $\pm$ 0.02 & 0.04\\
\hline
\end{tabular}
{\footnotesize  Relaxed clusters are defined in Section 6.1}
\label{tab:relation}
\end{table}

Radiative simulations are shown in green in Figure~1. In this case, a
direct comparison with the SPH results is less straightforward given
the differences in the subgrid model of supernova feedback
(kinetic for SPH simulation and thermal for AMR).
Nonetheless, there are a few general results we can glean from these
comparisons.  We find that while SPH and AMR respond similarly in the
{\it M} and {\it O} regions, they do contrast in the innermost part of
the clusters. The majority of the discrepancies between the mass-weighted
temperature and the spectroscopic-like one is generated in the core of
the AMR clusters (Figure~4) whereas in the same region the temperature variations for SPH
clusters are generally smaller and limited to the innermost radial
bins of hot systems (Figure~3).  The \tmw$-\Delta T$ relation for
\csfamr\ has a slope that is significantly steeper than
that of the \csfsph\ simulations (Figure~1).
Once the central 10\%
of $R_{200}$ ($\sim$15\% of $R_{500}$) is removed the amplitude
$\Delta_T$ of AMR simulations drops. Indeed, 
the majority of the systems show small temperature variations
$\Delta_T \lesssim 1$~keV of the ICM outside the core (bottom-right panel of Figure~4).

Dividing the set of simulated clusters in mass bins, we find that SPH
and AMR codes produce similar temperature variation ($\Delta_T \simeq
0.2-0.3$ keV at $R_{500}$ and $R_{200}$) for clusters with $M_{200}<
5\times 10^{14} $\minv, whereas the results differ considerably
($\Delta_T \simeq$ 0.7--1~keV for AMR and $\Delta_T \simeq$2.5~keV for
SPH) for more massive objects.

Finally, the AMR mixing being efficient both in nonradiative and
radiative simulations, we find that the spectroscopic-like determination of AMR clusters is
less influenced by the physics than for SPH simulations (left panel of
Figure~4 compared with left panel of Figure~2).

\begin{figure}[ht!]
\centering
\includegraphics[width=0.45\textwidth]{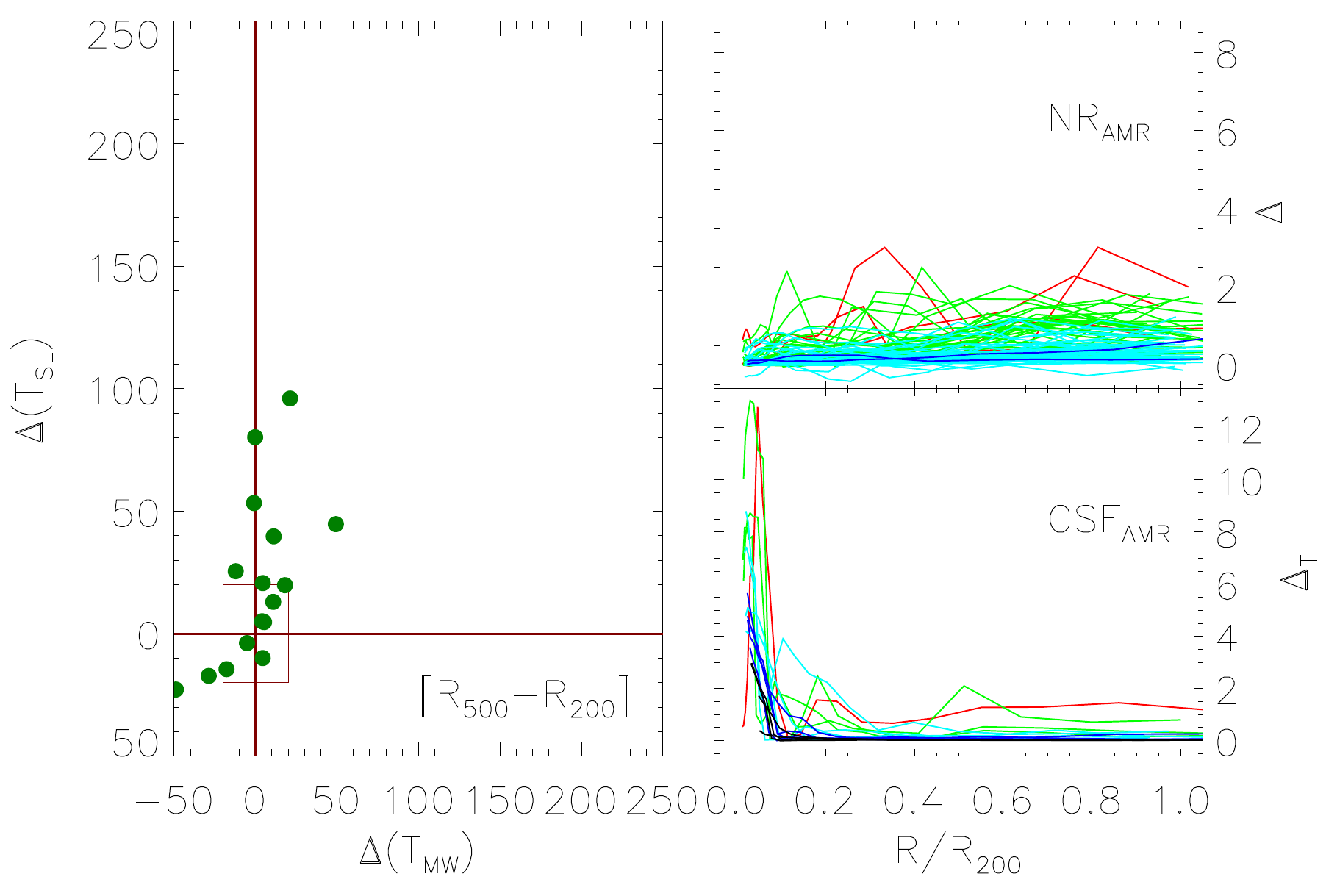}
\caption{Left panel: difference between the temperatures measured
  within AMR simulations of N07: the radiative case is compared to
  nonradiative one, similar to Figure~2. Right panel: profiles of
  temperature variation, $\Delta_T=$ \tmw-\tsl, for the different sets
  of simulated AMR clusters. Color code as in Figure~3.}
\end{figure}

\section{Characterization of thermal structures}

In the previous section, we show that the more efficient mixing of
mesh-based codes reduces the temperature variations of AMR clusters. At
the same time, by converting the low-entropy gas into stars through
the cooling and star formation processes, radiative simulations are
characterized by a less thermally perturbed ICM.
In this section, we evaluate how temperature inhomogeneities relate to
density perturbations.

\subsection{Log-normal Distributions}

The density, pressure, and temperature distributions of the simulated
ICM are approximately log-normal
\citep{rasia.etal.06,kawahara.etal.07,khedekar.etal.13,zhuravleva.etal.13}
with secondary peaks
in correspondence of subclumps.\footnote{The distributions of radiative
simulations produced by {\it all} gas elements is characterized by a
distinctive tail at high density or low temperature caused by the
overcooled dense blobs. However, after applying the cut described in
Section~2.3, this feature vanishes.}

We calculate the (decimal-base) logarithmic  gas density and temperature distributions
in logarithmically equispaced radial shells. We call $\rho_G$ and
$T_G$ the centers of the respective Gaussian distributions and $\sigma_{\rho}$
and $\sigma_{kT}$ their standard deviations.
\begin{figure}[ht!]
\centering
\includegraphics[width=0.5\textwidth]{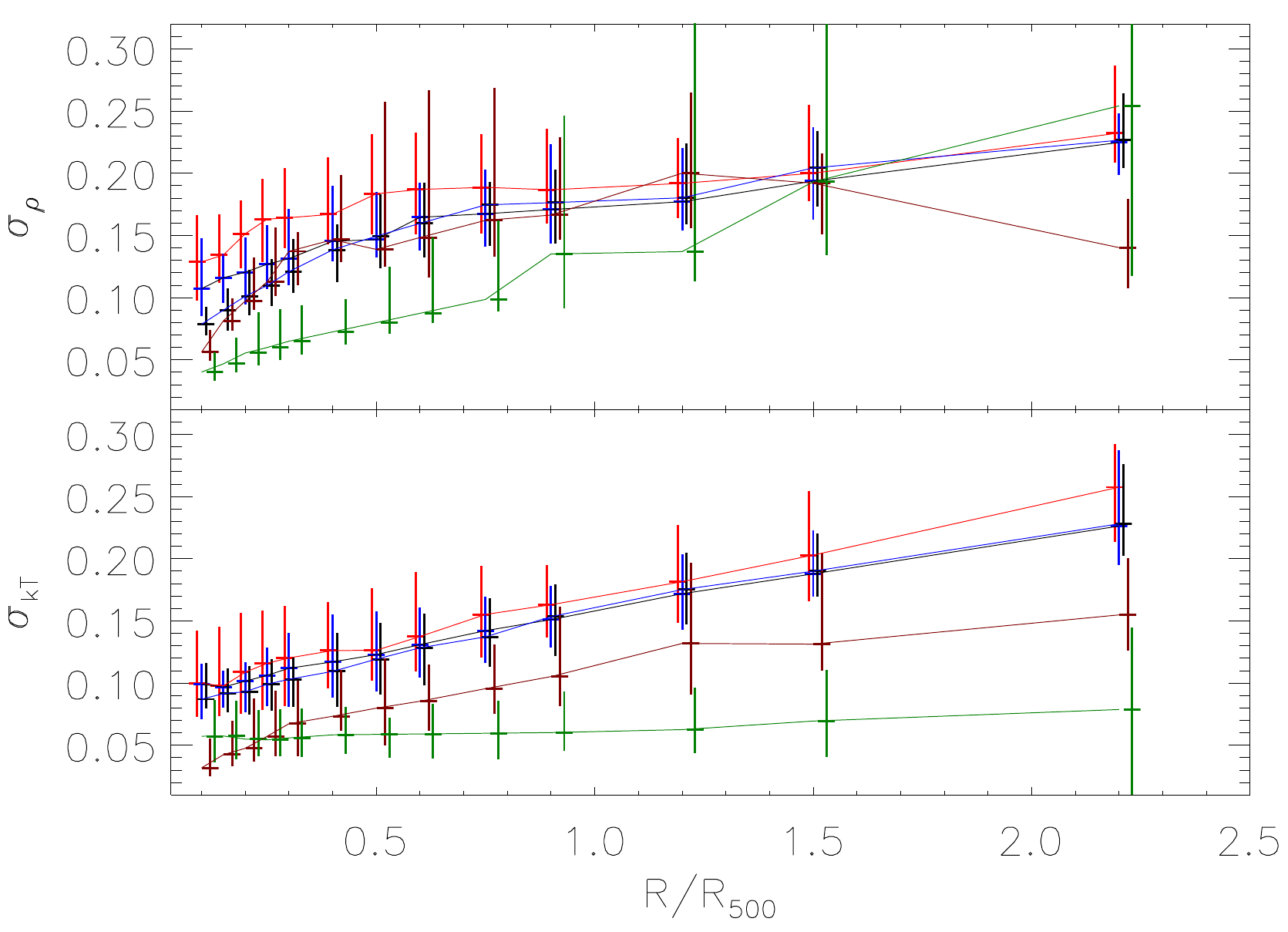}
\caption{ Radial profiles of the median widths of density and temperature
  log-normal distributions. Vertical bars span from the first to the third quartile. 
  Red,  blue, black, brown, and green refer, respectively, to \nrsph, \csfsph, \agnsph, \nramr, and \csfamr. }
\label{fig:siga}
\end{figure}
In Figure~\ref{fig:siga} we show the median radial profile of the
density and temperature dispersions. The temperature dispersion
profiles confirm the results outlined in Section~3. The density
dispersion profiles are close to one another, especially at a large
distance from the center. For $r>0.3\times R_{500}$, the
$\sigma_{\rho}$ profile of the \nramr simulations is consistent with
all profiles of the SPH set. For $r>0.7\times R_{500}$, the 
\csfamr\ also agree within the errors. 
In other words, the degree of substructures, which increases the width
of the gas density distribution, is comparable in the two
codes. Despite this, SPH clusters are characterized by a higher level
of temperature fluctuations. This suggests that the SPH temperature structure,
generated by the presence of dense clumps, is further
perturbed by other phenomena such as the persistence of the cold
stripped gas. 
This is particularly evident in the innermost
region where \nramr\ and \nrsph\ depart from one another. The reduced
density dispersion in the AMR simulations further proves the ability by the mesh-code
to disrupt infalling substructures, to quickly thermalize the stripped gas, and to maintain homogenous the cluster central regions.
%
%
\begin{figure}[ht!]
\centering
\includegraphics[width=0.5\textwidth]{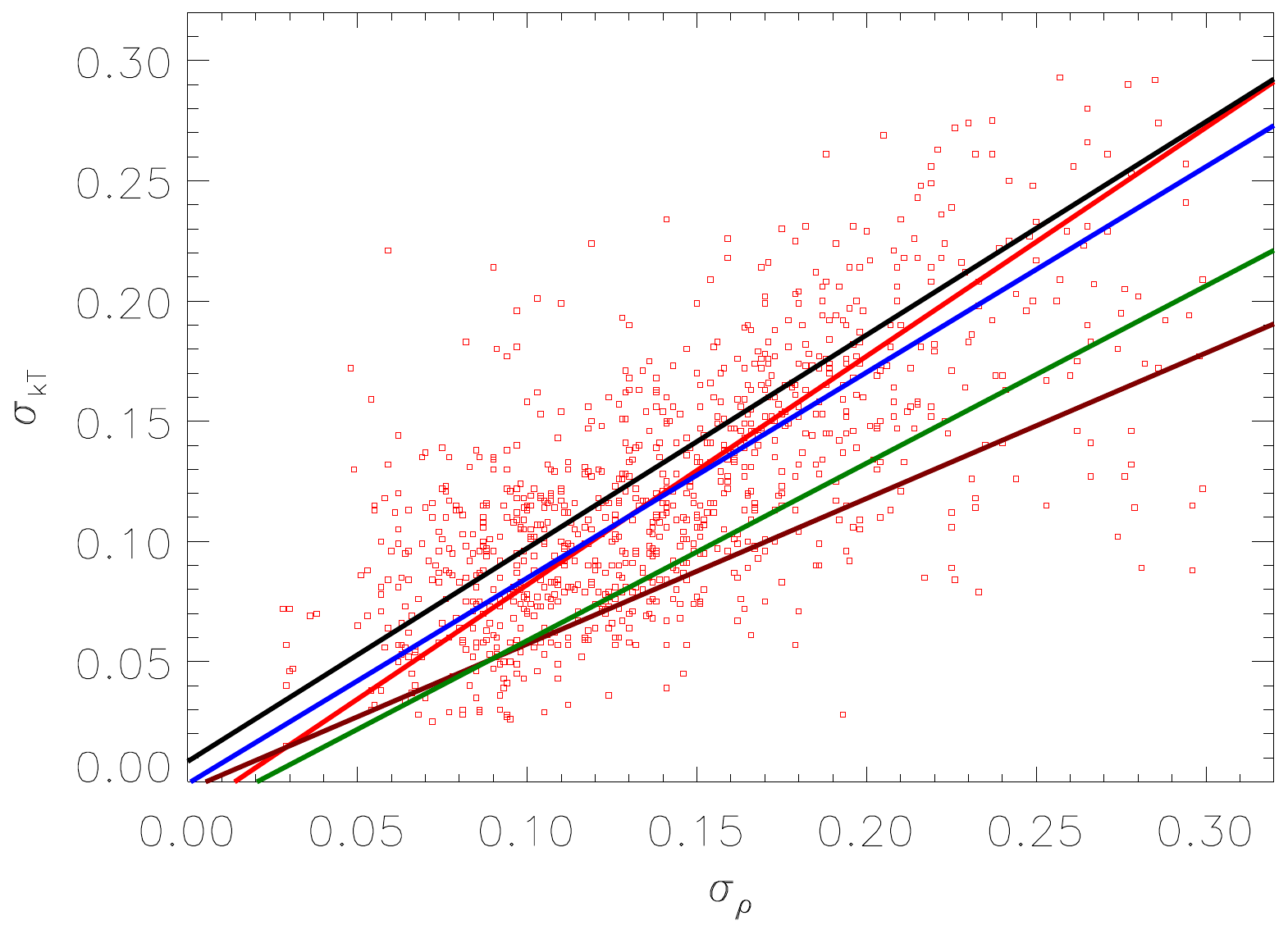}
\caption{Correspondence between $\sigma_{\rho}$ and $\sigma_{kT}$
  measured in each radial shell. For clarity, we plot the points only for \nramr\ and omit the  data points of the other physics
  whose best-fit linear relations (Equation (2)) are, however, shown with  the same color code as in Figure~5.}
\label{fig:corr}
\end{figure}

Another representation of this situation is presented in Figure~6 where we
plot the best-fit relations of the density dispersion versus the
temperature dispersion:
\begin{eqnarray}
& & {\rm for  \  \  NR_{SPH} :} \  \  \sigma_{kT}=0.95 \times \sigma_{\rho}  - 0.01; \nonumber \\
& & {\rm for  \  \  CSF_{SPH} :} \  \  \sigma_{kT}=0.85 \times \sigma_{\rho};  \nonumber \\
& & {\rm for  \  \  AGN_{SPH} :} \  \ \sigma_{kT}=0.89 \times \sigma_{\rho};\\
& & {\rm for  \  \  NR_{AMR} :} \  \ \sigma_{kT}=0.60 \times \sigma_{\rho}; \nonumber \\
& &  {\rm for  \  \  CSF_{AMR} :} \  \ \sigma_{kT}= 0.74 \times \sigma_{\rho} -0.01\nonumber.
\end{eqnarray}
The linear fits are derived using a bisector approach.  
At parity of density fluctuations, SPH clusters have higher temperature fluctuations.
For example, for $\sigma_{\rho} \approx 0.2$, the SPH temperature dispersion is
30\%--50\% above the value of AMR systems.

\subsection{Is the Cold Gas in pressure equilibrium?}

The connection between density and temperature dispersions 
is not enough to determine
whether the perturbations are isobaric. If we assume that
clusters have an onion structure and that the density
and temperature of each radial shell is equal to $\rho_G$ and $T_G$, we
find that the two quantities are highly correlated with a positive
Pearson correlation coefficient: $\xi_G \equiv \xi (\rho_G, T_G) =
$0.5--0.8.  The gas density decreases toward the outskirts as the
temperature does.  
If any fluctuation is 
completely isobaric, the presence
of subclumps will not change the pressure of the ICM. In this situation, the pressure
distribution within each shell should have a negligible dispersion.
\begin{figure}[ht!]
\centering \includegraphics[width=0.5\textwidth]{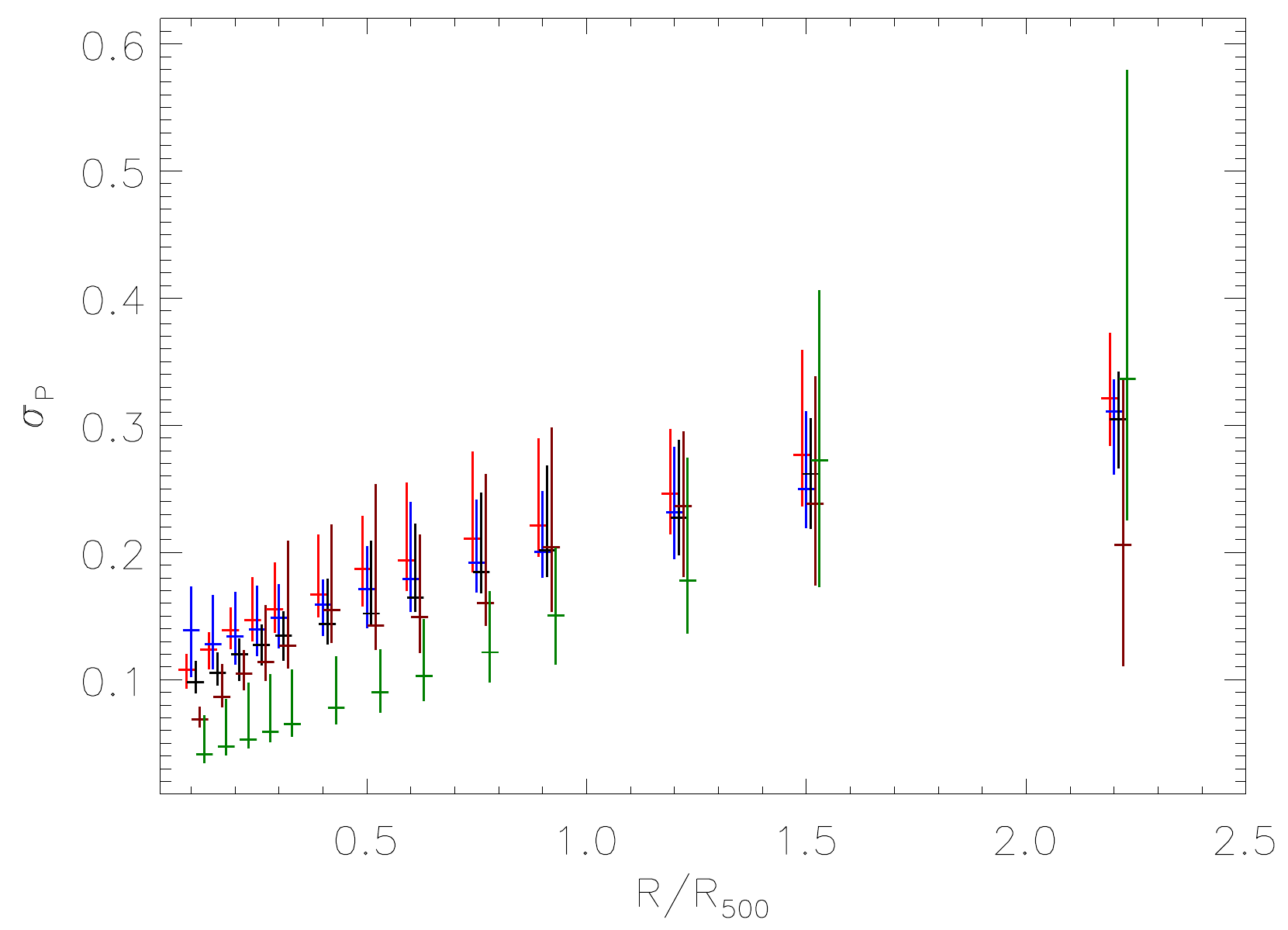}
\caption{Radial profile of the median widths of the pressure log-normal
  distributions. Vertical bars span from the first to the third
  quartile. The color code is that of Figure~5.}
\label{fig:sig_p}
\end{figure}

In Figure~7, we plot the pressure dispersion obtained from the log-normal fitting of the pressure distributions
extracted in the same radial bins as in Figure~5. By comparing the two figures, we notice that 
the pressure dispersion is even larger than the individual  density and temperature dispersions,
especially in the external radii.  \cite{zhuravleva.etal.13} explained this behavior by the increasing role with radius of sound waves and weak shocks as indicated by the ratio of the kinetic and thermal energies that changed from more isobaric in the core to more adiabatic farther away. 

The above test illustrates that the gas deviating from the average
behavior is not in pressure equilibrium. However, the test refers to all perturbations
with temperature higher as well as lower than the average.
We now focus only on
the cold gas because it is it is the responsible for the X-ray
temperature bias. For this purpose, we compute the correlation
coefficient between the density and the temperature of the 5\% coldest
gas in each of the regions {\it I}, {\it M}, and {\it O}.\footnote{As in the
rest of the paper, the cold-gas selection is done after the
application of the R12 cut.} The results are shown in Figure~8.
\begin{figure}[ht!]
\centering \includegraphics[width=0.4\textwidth]{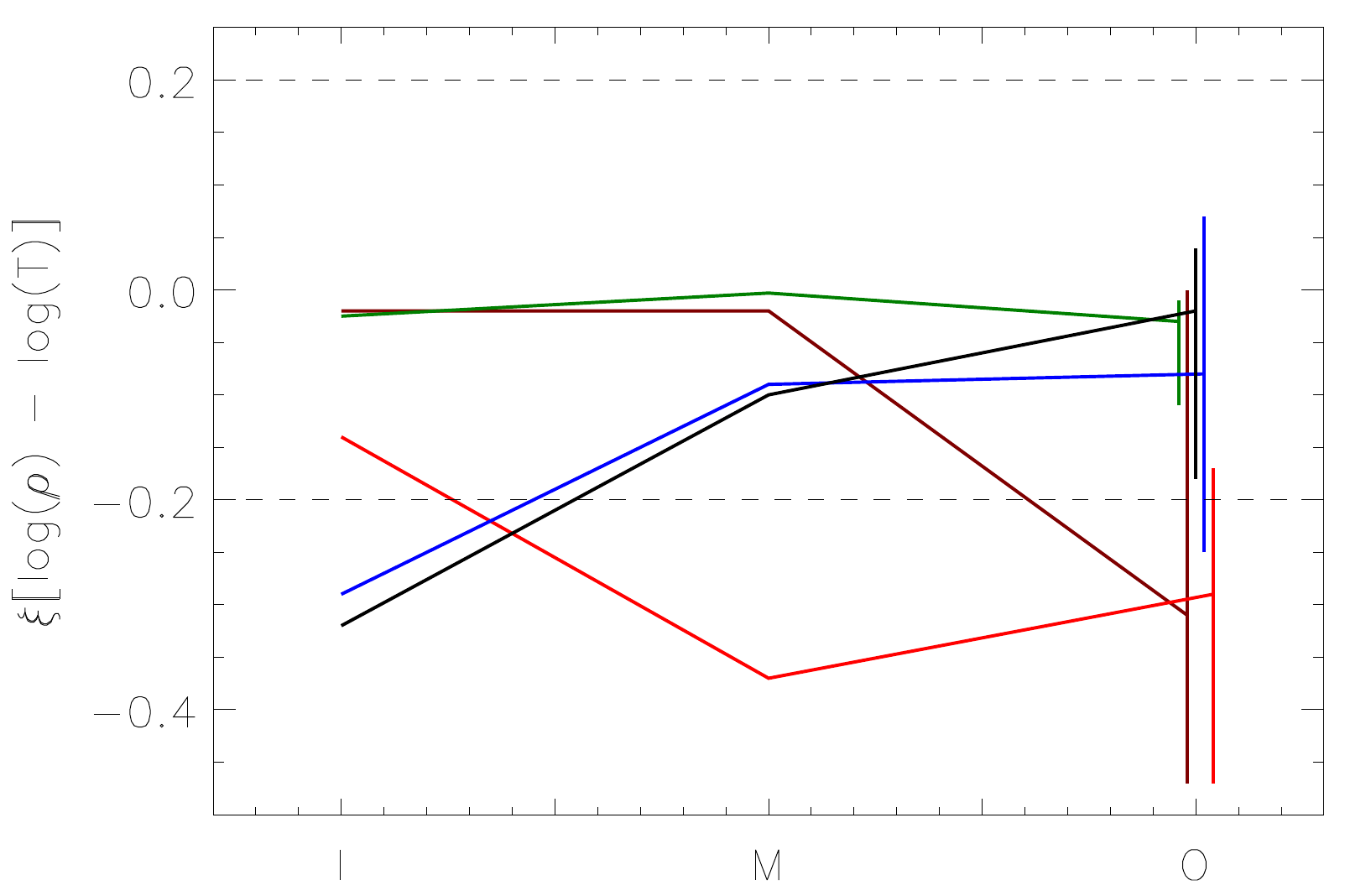}
\caption{Median of the correlation coefficients between the gas density and
  temperature of the 5\% coldest gas in each of the three regions {\it
    I}, {\it M}, and {\it O}. Only the median values of the are
  shown. For the {\it O} region, we also overplot the distance between
  the first and third quartiles as measure of the variance of the
  sample. The color code is the same of Figure~5.}
\end{figure}
%
In the majority of the simulations and regions, the values of $\xi$ are
between $-0.20$ and 0.20 indicating no correlation between the
temperature and the density of the coldest gas. Isobaric perturbations
are possible in the external regions of nonradiative simulations and
in the central region of the SPH radiative simulations. In the NR samples,
the coldest gas is likely associated with dense merging substructures. 
The survival time is longer in \nrsph\ simulations,
producing $\xi<-0.2$ even in the {\it M} region, while
the efficient mixing of \nramr\  reduces the presence of
cold clumps already at $R_{500}$. Moving inward. the amount of
dispersion in the temperature distribution decreases and the coldest
gas is no longer exclusively associated with clumps.  The negative $\xi$
in the inner region of SPH radiative clusters is, instead, caused by the
presence of a colder and denser core.  We
repeat the calculation of the correlation coefficient by accounting for the 10\% 
and the 25\% of the coldest gas in each region. The
qualitative tendency of the results holds. In conclusion, there is no evidence that
perturbations, and specifically cold inhomogeneities, are in pressure
equilibrium among them or with the diffuse medium.

\section{Comparison with Observations}
\subsection{Characterization of Density and Temperature Distribution.}

To compare with the results of \citet{frank.etal.13} we follow
their approach, create the emission-measure-temperature
distribution, and compute the median value, $T_{\rm med}$, and the
dispersion, $\sigma_{kT, EM}$:
\begin{equation}
\sigma_{kT, EM}=\sqrt{\frac{\sum_i [kT_i - \langle kT \rangle]\times EM_i}{\sum_i EM_i}}\,,
\label{eq:sigt}
\end{equation}
where $\langle kT \rangle =\Sigma_i (kT_i \times EM_i)/ \Sigma_i
(EM_i)$ is the mean of the emission-measure-weighted temperatures and
the emission measure is defined as $EM=m\times \rho$ \citep[see
  also][]{biffi.etal.12}.

  Figure~9 compares the width of the temperature distribution of \csfsph, \csfamr,
  and ${\rm TH.C_{SPH}}$ simulated clusters, calculated according to
  Equation (\ref{eq:sigt}), and the results of XMM--Newton observations by
  F13.  The temperature distribution analysis is carried out in the
  inner region ($r<R_{2500}$) of clusters. Results related to the nonradiative
  simulations or to the other regions are reported in Tables~3 and 4  for
  SPH and AMR, respectively.  For reference, we computed the best-fit
  linear relation following a Bayesian approach \citep{kelly.etal.07}
  to the sample analyzed by F13 selecting only objects with $T_{\rm
    med}<7.5$ keV. Simulations conducted by F13, indeed, showed that, in this range, the temperature
  derived via the smoothed-particle interference technique is trustable at
  better than 10\% while it is biased for hotter objects in part because the
  lack of sensitivity of XMM--Newton at higher energies (see Figure~9 of F13).
  The resulting
  relation is $\sigma_{kT}=0.60^{(\pm 0.2)} + 0.27^{(\pm 0.05)} \times
  kT_{\rm med}$ (black line and shaded gray area in Figure~9).

\begin{figure}[ht!]
\centering
\includegraphics[width=0.4\textwidth]{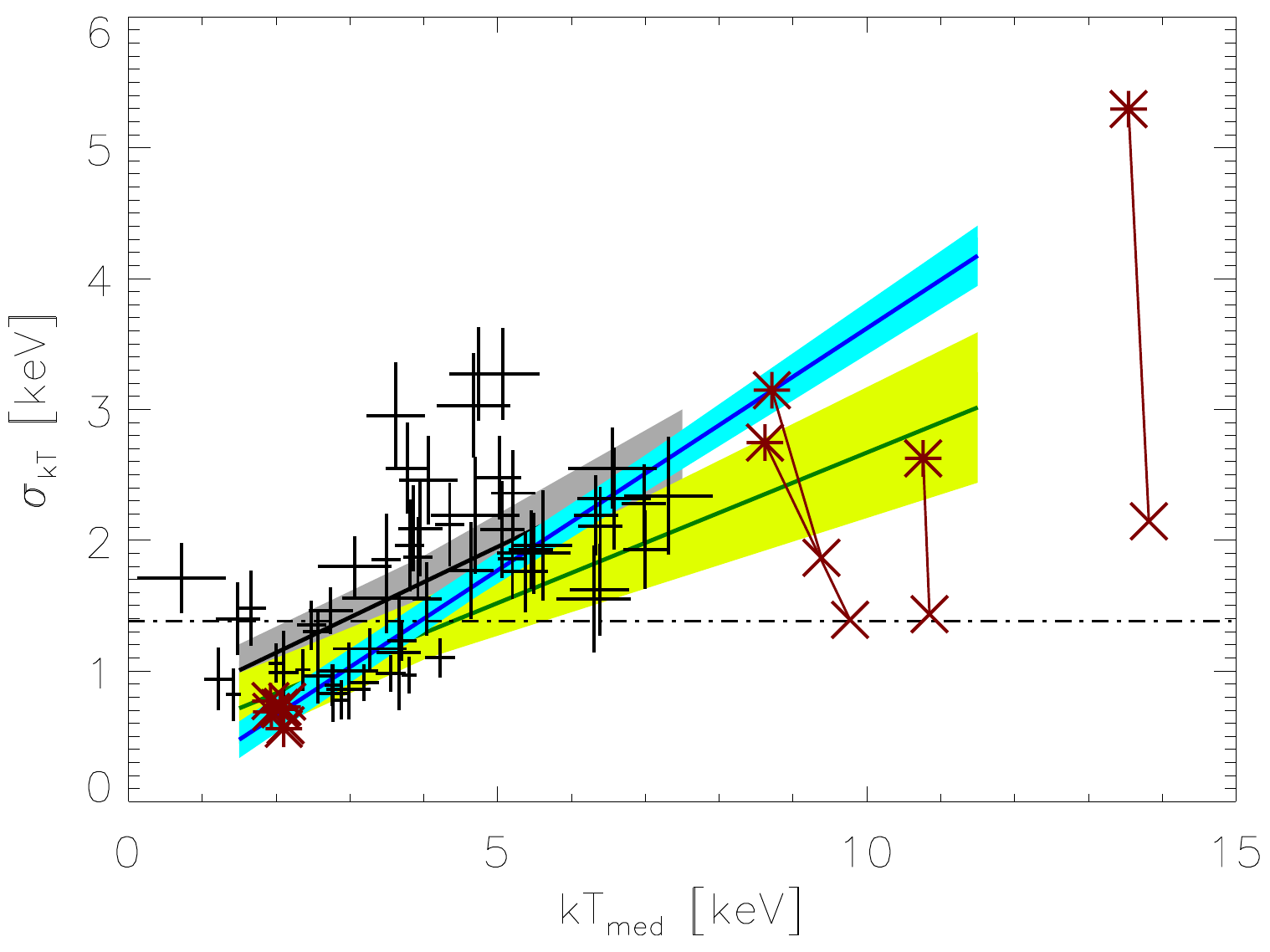}
\caption{Relation between width of the temperature distribution,
  $\sigma_{kT}$, and median cluster temperature for the {\it I}
  region.  F13 data points,  the
  corresponding best fit ($\sigma_{kT}=0.60{(\pm 0.2)} + 0.27{(\pm 0.05)} \times
  kT_{\rm med}$), and 1$\sigma$ uncertainty are given by the black
  crosses, black line, and gray region, respectively. The best-fit
  relations and uncertainties for the inner regions of simulated
  clusters are shown in blue for \csfsph\ (Table~4) and green for
  \csfamr(Table~5). Brown asterisks and crosses refer to the nine clusters simulated 
  with and without thermal conduction,
  respectively.}
\end{figure}

The comparison with numerical simulations demonstrates that CSF
simulations are in reasonable agreement with the results by F13 over
the temperature range probed by the current observations ($\simeq$
2--7 keV).  A word of caution, however, needs to be added because the
probed region is affected by modeling uncertainties.
Indeed, the origin of the temperature inhomogeneities in the {\it I}
region is different in AMR and in SPH: while for AMR it is mostly due to
the large temperature variation present in the cluster core
(Figure~4), for SPH simulations it is caused by the survival of substructures and their stripped gas
(Figure~3).  The remarkable influence of the core on AMR simulations
is clear by comparing Figures~9 and 10. The latter refers to the same {\it I} region
with the core (defined as $R<0.15\times R_{500}$) removed.
%
\begin{figure}[ht!]
\centering
\includegraphics[width=0.4\textwidth]{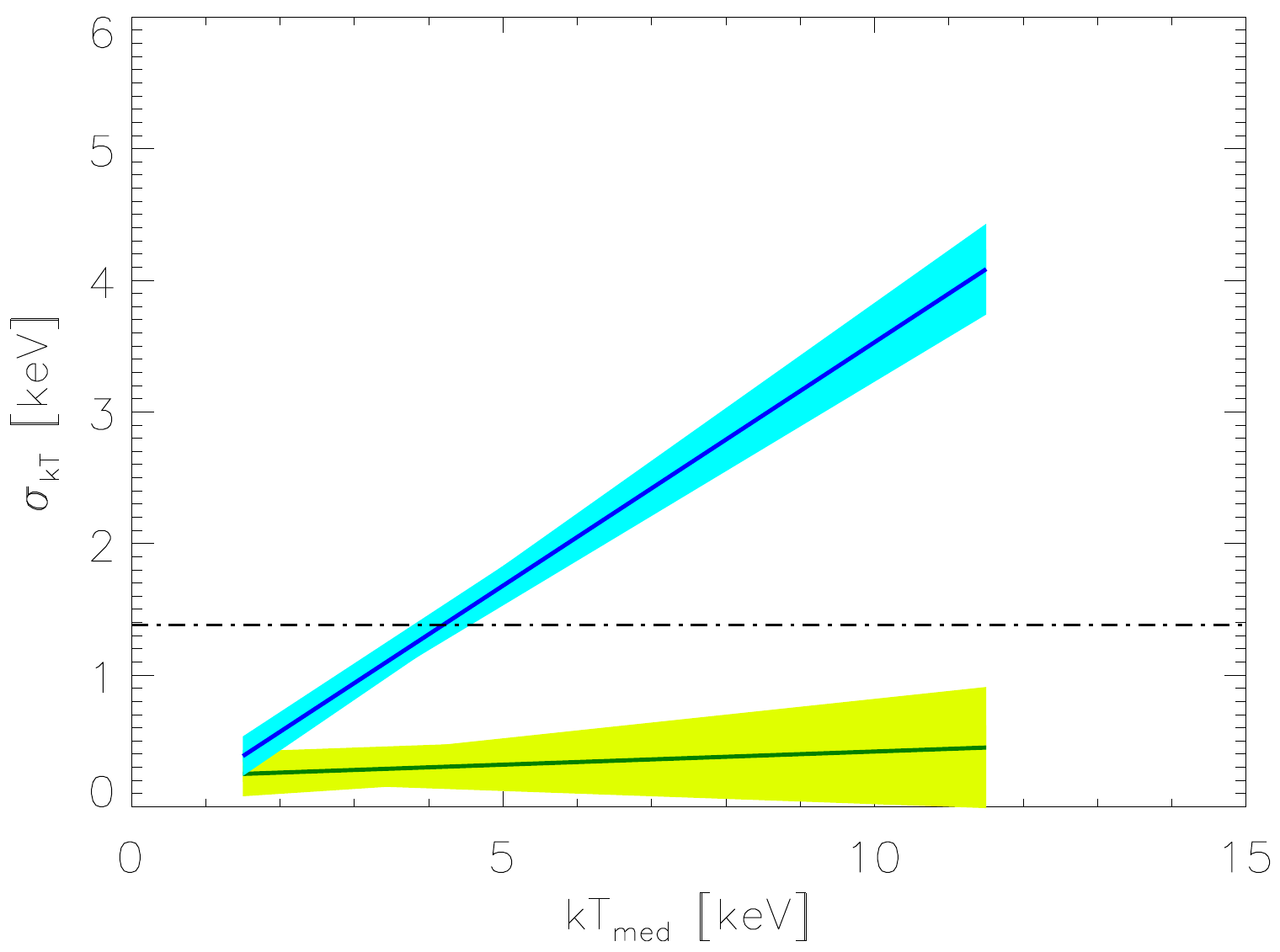}
\caption{Best-fit relations and uncertainties of the {\it I} regions of
 \csfsph\  clusters (blue) and  \csfamr\ objects (green) once the central $0.15 \times
  R_{500}$ are excised.}
\end{figure}
%
Including AGN feedback in SPH simulations slightly reduces the
width of the temperature distribution, because AGNs expel gas
from the substructures decreasing the ICM temperature
inhomogeneity. The resulting $\sigma_{kT}-T$ relation becomes very
similar to the observed one (see Table~3).

Thermal conduction reduces the values of $\sigma_{kT}$, especially at
high temperature, where conductivity becomes efficient
\citep{dolag.etal.04}. The corresponding $\sigma_{kT}$--$T$ relation
is shallower than the extrapolation of the observed one, suggesting
that measurements of the ICM temperature distribution of very hot
clusters (currently not available) can be used to constrain the degree of thermal conduction in
the intracluster plasma.

\begin{table}[ht!]
\caption{SPH set: best-fit parameters, their 1 $\sigma$ errors, and scatter of
  the linear relation: $\sigma_{kT}=A+B \times k T_{\rm med}$}
\centering
\begin{tabular}{|lccc|} 
\hline
\multicolumn{4}{|c|}{all clusters}\\
 & A$\pm$ err(A) &B $\pm$ err(B) & scatter \\
\hline
\nrsph, I              &  -0.32$  \pm$ 0.19  & 0.43  $\pm$ 0.03  & 0.58\\
\nrsph, M            &  -0.36$ \pm$  0.13  & 0.63  $\pm$ 0.03  & 0.38\\
\nrsph, O            &  -0.20  $\pm$ 0.14  & 0.64  $\pm$ 0.04 &  0.40\\
\hline
\csfsph, I             &   -0.15$  \pm$ 0.24 & 0.38  $\pm$ 0.04  & 0.67\\
\csfsph, M           &  -0.46$ \pm$ 0.18  & 0.55  $\pm$ 0.04 & 0.51\\
\csfsph, O           &  -0.37 $ \pm$ 0.12 & 0.68  $\pm$ 0.03 &  0.35\\
\hline
\agnsph, I           &   0.05 $ \pm$ 0.17  &  0.34   $ \pm$  0.03 & 0.47 \\
\agnsph,M          &  -0.37 $ \pm$ 0.16  &  0.51  $ \pm$ 0.03 & 0.46 \\
\agnsph, O         &  -0.41 $ \pm$ 0.11  &  0.70   $ \pm$ 0.03 & 0.31 \\
\hline
${\rm TH.C_{SPH}}$, I   & 0.26   $ \pm$ 0.09  & 0.13    $ \pm$ 0.01 & 0.05 \\ 
${\rm TH.C_{SPH}}$, M &  -0.13 $ \pm$ 0.07  &  0.29   $ \pm$ 0.01 & 0.04 \\
${\rm TH.C_{SPH}}$, O &  -0.06  $ \pm$  0.02 &   0.27  $ \pm$ 0.01 &  0.02\\
\hline
\hline
\multicolumn{4}{|c|}{only relaxed}\\
 & A$\pm$ err(A) &B $\pm$ err(B) & scatter \\
\hline
\nrsph, I              &  -0.45$  \pm$ 0.20  & 0.34  $\pm$ 0.04  & 0.29\\
\nrsph, M            &  -0.60$ \pm$  0.20  & 0.63  $\pm$ 0.04  & 0.28\\
\nrsph, O            &  -0.29  $\pm$ 0.09  & 0.66  $\pm$ 0.03 &  0.12\\
\hline
\csfsph, I             &   0.24$  \pm$ 0.25 & 0.21  $\pm$ 0.04  & 0.33\\
\csfsph, M           &  -0.30$ \pm$ 0.21  & 0.41  $\pm$ 0.04  & 0.29\\
\csfsph, O           &  -0.71 $ \pm$ 0.09 & 0.75  $\pm$ 0.03 &  0.12\\
\hline
\agnsph, I           &   0.41 $ \pm$ 0.15  &  0.22   $ \pm$  0.02 & 0.20 \\
\agnsph,M          &  -0.34 $ \pm$ 0.14  &  0.40  $ \pm$ 0.03 & 0.19 \\
\agnsph, O         &  -0.57 $ \pm$ 0.10  &  0.69   $ \pm$ 0.03 & 0.12 \\
\hline
\end{tabular}
\label{tab:relation}
{\footnotesize  Relaxed clusters are defined in Section 6.1}
\end{table}

\begin{table}[ht!]
\caption{AMR set: best-fit parameters, their 1 $\sigma$ errors, and scatter of
  the linear relation: $\sigma_{kT}=A+B \times k T_{\rm med}$}
\centering
\begin{tabular}{|lccc|} 
\hline
\multicolumn{4}{|c|}{all clusters}\\
 & A$\pm$ err(A) &B $\pm$ err(B) & scatter \\
\hline
\nramr, I               &   0.04$  \pm$ 0.10   & 0.23  $\pm$ 0.02  & 0.22\\
\nramr, M             &  -0.19$  \pm$ 0.12   & 0.38  $\pm$ 0.04  & 0.24 \\
\nramr, O             & -0.12  $ \pm$ 0.12   & 0.50  $\pm$  0.05 &  0.26 \\
\hline
\csfamr, I               &   0.37  $ \pm$ 0.27  &  0.23 $\pm$ 0.05  & 0.51  \\
\csfamr, M             &   -0.17 $ \pm$ 0.09  &  0.32 $\pm$ 0.02  & 0.18 \\
\csfamr, O             &   -0.40 $ \pm$ 0.18 &   0.59$\pm$   0.07&  0.40\\
\hline

\multicolumn{4}{|c|}{only relaxed}\\
 & A$\pm$ err(A) &B $\pm$ err(B) & scatter \\
\hline
\nramr, I               &   0.21$  \pm$ 0.07   & 0.17  $\pm$ 0.02  & 0.10\\
\nramr, M             &  -0.03$  \pm$ 0.14   & 0.24  $\pm$ 0.06  & 0.20 \\
\nramr, O             &  0.15  $ \pm$ 0.13  & 0.25  $\pm$  0.08 &  0.18\\
\hline
\csfamr, I               &   0.61  $ \pm$ 0.32  &  0.17 $\pm$ 0.06  & 0.38  \\
\csfamr, M             &   0.01 $ \pm$ 0.10  &  0.22 $\pm$ 0.03  & 0.11 \\
\csfamr, O             &   0.08 $ \pm$ 0.09 &   0.22$\pm$   0.05&  0.11 \\
\hline
\end{tabular}
\label{tab:relation}
{\footnotesize  Relaxed clusters are defined in Section 6.1}
\end{table}


\section{Consequences for the X-ray mass}

Using the gas as a tracer and assuming HE, the
total mass of a system within a certain radius $r$ is calculated as
\begin{equation}
M(<r) = - \frac{r T(r)}{\mu m_p G} \times \left[ \frac{d \log \rho_{gas}}{d \log r} + \frac{d \log T}{d \log r} \right],
\end{equation}
where the temperature and the derivatives are computed at the radius $r$, $\mu \approx 0.59 $ is the mean molecular weight, $m_p$ is the proton mass, and $G$ is the gravitational constant.
  Several works based on simulations already pointed out that on
top of the thermal pressure another term is needed to counterbalance
the gravity accounting for 10\%--15\%
of the total mass \citep[see][for a review on X-ray mass
measurement]{ettori.etal.13}.  In this section, using the simulations with radiative physics, we estimate the
potential extra contribution generated by the X-ray temperature bias
\citep[][R12]{rasia.etal.06}.  For this purpose, we derive the 
mass  by adopting both the mass-weighted temperature and
the spectroscopic-like one. We call $M_{MW}$ and $M_{SL}$ the
respective masses at $R_{500}$.
\begin{figure*}[ht!]
\centering
\includegraphics[width=0.4\textwidth]{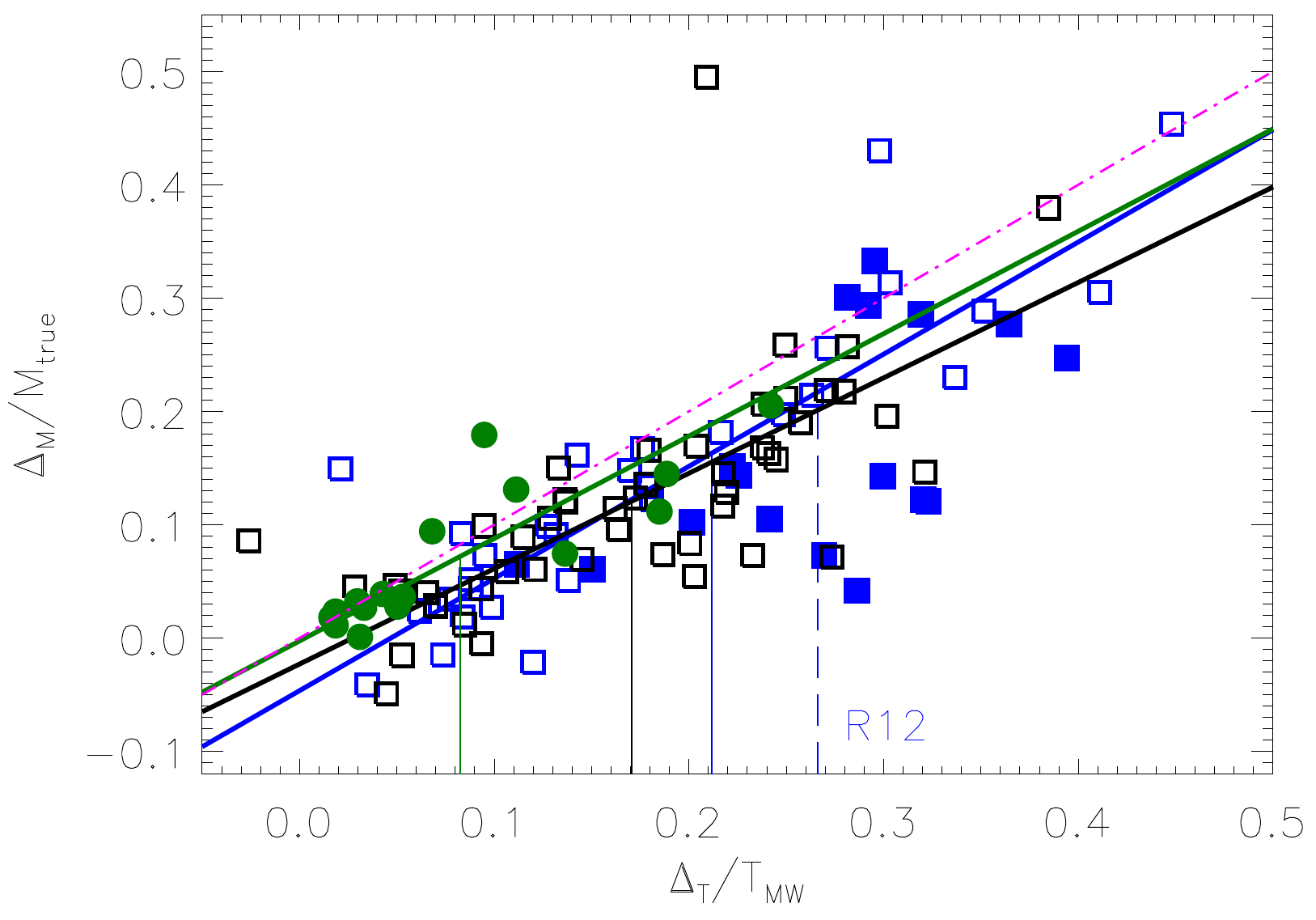}
\includegraphics[width=0.4\textwidth]{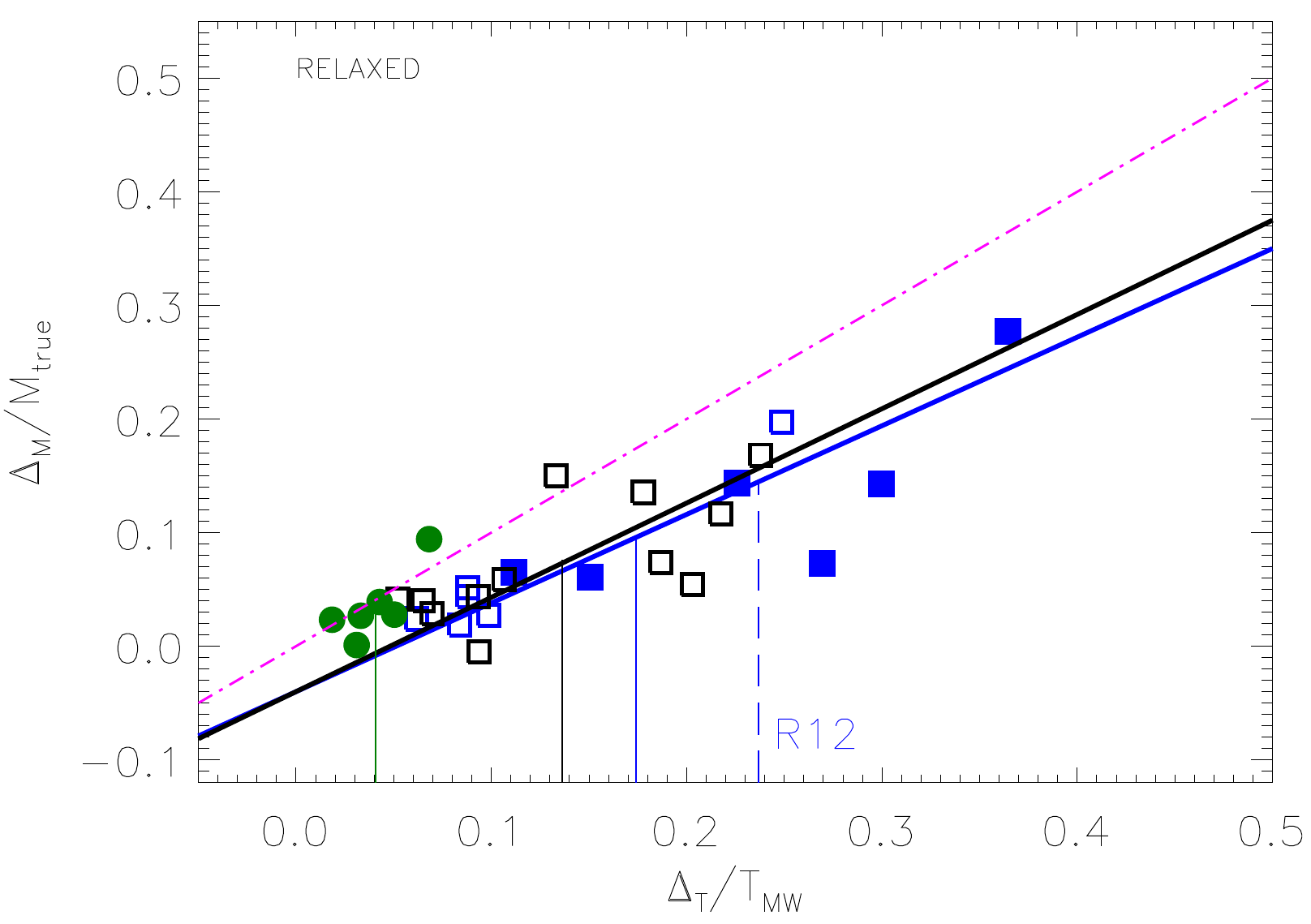}
\caption{Mass bias, $(\Delta _M=M_{MW}-M_{SL})/M_{\rm true}$, versus
  temperature bias, $(\Delta_T= T_{MW} - T_{SL})/T_{MW}$ at $R_{500}$.  The entire
  samples are shown in the left panel, and only relaxed objects are isolated in
  the right panel.  Green circles and lines refer to \csfamr. Blue and
  black squares and lines show \csfsph\ and \agnsph, respectively.
  The magenta line represents the identity relation. The median
  values of the temperature bias are plotted as vertical lines.  The
  rightmost dashed blue line corresponds to the R12 sample identified by the
  filled blue squares. The bisector best fit relations in the form
  $\Delta_M/M_{\rm true}=A+B \times \Delta_T/T_{MW}$ have parameters
  $(A;B)=(0;0.9)$ for \csfamr; $(A;B)=(-0.07;1.07)$ for \csfsph\ and
  $(A;B)=(-0.04;0.91)$ for \agnsph. The AGN relation changes only
  slightly for relaxed objects while the slope of  \csfsph decreases
  to 0.8.}
\end{figure*}

The relation between the temperature variation normalized to the
mass-weighted temperature and the normalized mass variation is shown
in  Figure~11.  Once again, while minor differences are detected
between the two SPH feedback mechanisms (\csfsph\ versus \agnsph), we
notice a separation between the normalization of SPH and AMR
simulations: at fixed $\Delta_T$\tmw\ AMR clusters have a larger $\Delta_M/M_{\rm true}$ associated with them.
The explanation relies on the fact that around $R_{500}$, the
$T_{SL}$ profile is steeper than the $T_{MW}$ profile for SPH
simulated clusters.  The
temperature derivative, within the $M_{SL}$ expression, is, thus, more
negative and as such the mass bias is effectively reduced.  With
vertical lines, we report the median values of the relative
temperature variations of the radiative samples. The mass
bias of the R12 sample is, on average, the most affected by
temperature inhomogeneities because the sample contains massive
systems that are experiencing several merging
events. The set of M10 is not shown in the figure, however, the effect
of thermal conduction is such to locate all nine clusters in the same
position in the plane: $\Delta_T/$\tmw\ $\sim \Delta_M/M_{\rm true} < 0.12$.

As a final step, we study how the points move in the mass $-$ temperature plane according to
the two temperature definitions. We derive the linear fit in the form $\log (M)=N + \alpha \times \log(T)$ for the following relations: $M_{\rm true}-T_{\rm MW}$, $M_{\rm MW}-T_{\rm MW}$, and $M_{\rm SL}-T_{\rm SL}$. The power-law index $\alpha \approx 1.5-1.6$ is always consistent within 1 $\sigma$ error among the three relations for all radiative simulations. The normalizations, $10^N$, of the second and third relations vary with respect to the first case according to the median ratios $M_{\rm MW}/M_{\rm true}$ and $M_{\rm SL}/M_{\rm true}$. Respectively, these are equal to 15\% and  20\% in SPH and they are about 5\% in AMR.

\subsection{Relaxed sample}
In this section, we restrict the study of the mass bias to relaxed
systems.
Simulations show that
the degree of inhomogeneities in the medium depends on the dynamical
state of the cluster.  For example, recently, \cite{vazza.etal.13}
showed that the baryon fraction can be twice as biased in perturbed
systems. At the same time, \citet{zhuravleva.etal.13} demonstrated
that the gas density distribution of unrelaxed clusters is 
higher with respect to relaxed clusters in a large interval of radii
([$R_{2500} \div R_{200}$]) and that the peaks of the
distributions of relaxed and perturbed systems have a significant separation. 

In the radiative sample, we define the relaxation of a cluster on the
basis of the X-ray morphology.  For the \csfamr\ set, we adopt the
classification of \cite{khedekar.etal.13} and
\cite{zhuravleva.etal.13} where 6 objects out of 16 are visually
recognized as X-ray regular. For the SPH samples, we measure the global
X-ray morphological parameter, $M_{\rm par}$
\citep{rasia.etal.12b,meneghetti.etal.14}, and impose $M_{\rm par}
<-1$. In the Appendix, this selection method is compared with the
mass-accretion-history parameter $\Gamma$ \citep[e.g.][]{diemer.etal.13}.  The
relaxed samples of SPH are composed by the 12 objects that satisfy
the condition in all of the three physics.

The relations analyzed in Sections~3 and 5 are rederived
 considering regular systems (see bottom part of all tables).
 The results for SPH simulations change slightly, often favoring a smaller normalization. In most of the cases, the relations of the entire sample and those of relaxed objects are consistent within 1$\sigma$ whenever the \tmw\ of the clusters is $T_{\rm MW} \leq 5$ keV whereas massive perturbed objects have higher temperature variation $\Delta_T$ and temperature dispersion $\sigma_{kT}$.
 The \tmw$-\Delta_T$ normalization and slope of the AMR relaxed systems change more drastically being always consistent with zero (with the exception of the central region). For each physics and region, the \tmw$-\sigma_{kT}$ relation is also shallower. At fixed \tmw\ the value of $\sigma_{kT}$ is lower by 10\% for \tmw$\leq 2$ keV, $20\%-25\%$ for \tmw$= 3-4$ keV, and $25\%-30\%$ for \tmw$\geq 5$ keV.

In the right panel of  Figure~11 we plot the influence of the
temperature variation on the mass bias for relaxed samples.
As expected, the \csfamr\ systems are distinguished by a low degree of
both temperature and mass variation. 
In this case, for the reduced range of both axes, we are not able to 
 linearly fit the points. 
For \csfsph\ simulations, the slope of the $\Delta_M/M_{\rm true}-\Delta_T/$\tmw\ relation decreases by 20\% with respect to the entire sample. The mean temperature variation of relaxed objects corresponds to a mass bias that is about half the mass
bias of the total sample.

\section{Conclusions}

Motivated by the discrepant results by N07, M10, and R12 on the HE
mass bias, we evaluate the degrees of temperature inhomogeneities
present in their simulated sets.  Structures in the ICM temperature
distribution are, indeed, the main sources of systematic bias in the
X-ray spectroscopic temperature measurement with direct consequences
for the HE mass estimate of X--ray clusters.  We analyzed four
different samples simulated with either {\tt {GADGET}}, an SPH code, or  {\tt ART}, an AMR algorithm.  The simulations implement various
prescriptions for the baryonic physics, including nonradiative gas and
processes of cooling, star formation and feedback by supernovae
or AGNs. A small sample of nine objects allowed us to
study the effect of thermal conduction.  After comparing
the degree of temperature structure and studying its nature, we tested
the predictions against the observational results of \cite{frank.etal.13} and
derived our conclusions on the consequences of the X-ray mass bias.  Our
main results can be summarized as follows:

\begin{itemize}
\item AMR simulations with nonradiative physics predict a lower
  degree of ICM temperature inhomogeneities with respect to SPH
  because the more efficient mixing destroys substructures
  during their infall within the cluster and quickly thermalizes the
  stripped gas.
 
\item The effect of baryonic physics in radiative
  simulations substantially reduces the differences between AMR and
  SPH simulations.  Radiative cooling removes cold and dense gas from its
  diffuse state, thus reducing the entropy contrast of the
  ICM. However, the discrepancies between the simulated sets are still significant
  at small radii ($R<R_{2500}$) mostly because of the complex physics
  of the core and the different implementations of the stellar feedback. 
  \csfsph~ and \agnsph~ simulations show similar response to
  temperature variations even if there is a systematic tendency to
  have less inhomogeneity in the presence of AGN. The inclusion of kinetic feedback 
  in the AGN model might, however, increase this difference. Thermal conduction 
  drastically smooths temperature variations and homogenizes  
  the ICM.

\item AMR and SPH produce a comparable amount of density
  inhomogeneities, especially in the nonradiative case and in the
  external regions (outside $\sim 0.7 \times R_{500}$). However, a
  fixed amount of density inhomogeneities presents a higher
  degree of temperature perturbations in SPH clusters.

\item The cold gas of nonradiative simulations is associated with
  dense clumps mostly connected to merging substructures. The
  radiative simulations instead present a negligible correlation between the
  temperature and density. This confirms the idea that the coldest gas is
  not in pressure equilibrium with the diffuse gas.

\item The emission-measure temperature dispersions of radiative simulations
  carried out by both codes match equally well the observational
  data of F13 even if for different reasons: the dispersion of AMR clusters 
  depends on the core physics while that of SPH is caused by
  the survival of substructures and the cold stripped gas. \\ 
    From an observational point of view, 
  more insights on the ICM processes might be provided
  by masking the core of observed clusters. 
 In this case, AMR clusters show a temperature dispersion
  consistent with zero over the entire temperature range while the $kT_{med}-\sigma_{kt}$ relation of SPH
  systems does not change significantly.
  Another solution could be to measure temperature variation at distances
  larger than $R_{2500}$. Indeed, the predicted dispersion grows rapidly with
  the radius: in the {\it M} region the difference in 
  $\sigma_{kT}$ between AMR and SPH increases by at least $\sim$
  60\% for all systems with temperature $T[R_{2500} - R_{500}] > 2$
  keV.
 Finally, the difference between AMR and SPH becomes more evident for
  high--temperature clusters.  
  For example, the difference between the predicted
  AMR and SPH dispersions is 16\% for a 5 keV
  cluster and grows up to 30\% for an 8 keV one.

\item The consequences for the X-ray mass bias caused by thermal
  fluctuations are similar among the radiative simulations. However, because
  the temperature variations are smaller in AMR simulations, their 
  mass bias can be a factor of two lower. The difference is even more
  marked when the sample of N07 is compared to R12 because the latter has
  massive objects with heavily disturbed ICM.
\item As expected, relaxed objects 
present  lower degrees of inhomogeneities, especially for AMR simulations. 
\end{itemize}

The exact determination of the temperature bias is sought because its
contribution to the X--ray mass bias might be as high as non-thermal
pressure support associated with ICM bulk motions (R12, Planck
Collaboration 2013).  Upcoming high--resolution X--ray spectroscopic
observations, e.g., with {\em ASTRO-H}, will help characterize gas
motions with direct implications for the mass calibration of clusters.
At the same time, detailed X--ray observations would be necessary to
extend the current description of ICM thermal fluctuation to larger
radii and including hotter systems. While pushing the capabilities of
the current generation of instruments to their limits will be
beneficial, a leap forward in these studies will be reached with the
advent of a next generation of high--sensitivity X--ray telescopes
such as the Athena+ X-ray observatory
\citep{nandra.etal.13,pointecouteau.etal.13}.

\acknowledgments

We are greatly indebted to Dunja Fabjan, Annalisa Bonafede, 
 and Luca Tornatore who produced the simulations used in this
work.   We thank Madhura Killedar for providing the offset measurements 
of Killedar et al. (2012)and  Eugene Churazov, Irina Zhuravleva, and the referee for their useful comments. 
We acknowledge financial support from: NSF AST-1210973; SAO
TM3-14008X (issued under NASA Contract No. NAS8-03060); CXC GO2-13153X
and HST GO-12757.01A; DFG Cluster of Excellence ``Origin and Structure
of the Universe", NSF AST-100981, NASA-ATP NNX11AE07G, NASA Chandra
Theory grant GO213004B, Research Corporation, Yale University, NSF
Graduate Student Research Fellowship; Alan D. Bromley Fellowship;
PRIN-MIUR 2012 grant ``Evolution of Cosmic Baryons"; PRIN-INAF 2012 grant; INFN FP7 Marie Curie Initial Training Network CosmoComp (PITN-GA-2009-238356).

\appendix
\section{Selection of relaxed cluster.}

Using the \nrsph, \csfsph, and \agnsph\ sets, we compare two approaches to select relaxed
systems: the first, more theoretical, uses the mass-accretion parameter,
$\Gamma$ \cite[e.g.][]{vazza.etal.13,diemer&kravtsov14}, and the second,
observationally oriented, considers the global morphological parameter,
$M_{par} $\citep{rasia.etal.12b}.

{\bf Dynamical state.}
The mass-accretion-rate parameter is a measure of the mass increase of
an object with time: 
\beq 
\Gamma=\frac{\log  M_{500}(z_2)-\log M_{500}(z_1)}{\log (1+z_1)-\log(1+z_2)} 
  \eeq
where the mass at redshift $z_2$ refers to the most massive progenitor
of the cluster at redshift $z_1<z_2$.  
The redshift of reference, $z_1$, corresponds to the
one used in this work and it is set equal to zero while $z_2$ is fixed to
0.25.  The redshift difference corresponds to 3 Gyr: a sufficient time to allow a substructure that merged before or
around $z_2$ to be completely incorporated into the main cluster but not
enough time to allow a substructure that merged afterwards
 to relax \citep{nelson.etal.14}.  The values of the $\Gamma$
parameters are not influenced by the ICM physics.  
We consider  $\Gamma=2$ as
the threshold  to distinguish between relaxed and
perturbed objects.  This factor corresponds to a mass
increase of about 35\% between redshift $z_2$ and $z_1$, equal to
the factor attributable only to the pseudo-evolution of clusters
\citep{diemer.etal.13}.
%

{\bf X-ray regularity: morphological parameters.}
The X-ray regularity is estimated through the global morphological parameter, $M_{par}$, defined as
  \beq
  M_{par}=\sum{\frac{X - < X >}{\sigma_X}}
\end{equation}
where $X$ represents an ensemble of morphological parameters, <>
denotes the mean values of the distribution of each parameter, and
$\sigma$ their standard deviation. The morphological estimators used
are: the centroid shift, $w$ \citep{mohr.etal.93}; the ellipticity,
$\epsilon$; the X-ray surface brightness concentration
\citep{cassano.etal.10}; and the third and fourth power ratios, $P_3$
and $P_4$ \citep{buote&tsai95}, such that $X \equiv [\log w, \log 1/c,
\log P_3, \log P_4, \epsilon]$\footnote{The presence of the logarithm
  is justified by the log-normal nature of the distributions of all
  morphological parameters with the exception of the Gaussian shape of
  the ellipticity distribution.}.  The means and standard deviations
of our \nrsph\ samples are comparable with those derived by
\citet{meneghetti.etal.14} who analyzed a much larger sample 
 taken from the MUSIC simulations
\citep{sembolini.etal.13}. Objects with $M_{par}$ below zero are by 
definition more regular than the average. To be more restrictive we
impose the limit of $M_{par} < -1$.

\begin{figure}[ht!]
\centering
\includegraphics[width=0.4\textwidth]{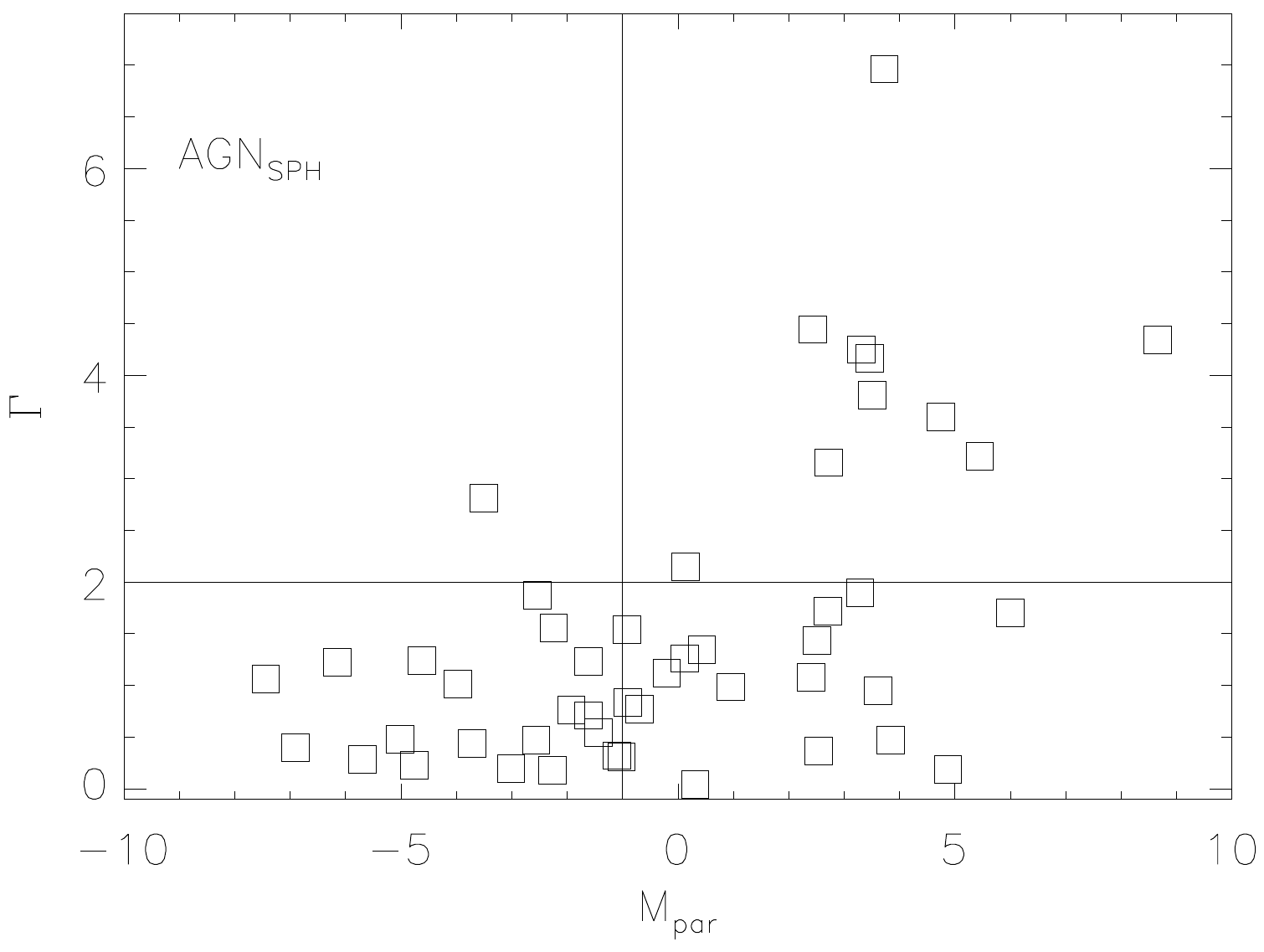}
\caption{Relation between mass-accretion-history parameter, $\Gamma$, and the morphological
  parameter, $M_{\rm par}$. Horizontal and vertical lines show the
  limits used to distinguish between dynamically relaxed and unrelaxed objects ($\Gamma = 2$) and between X-ray regular and disturbed X-ray images ($M_{\rm par}=-1$).}
\label{fig:fig11}
\end{figure}
Interestingly, $\Gamma$ shows a good degree  of
correlation with the X-ray morphological parameter $M_{\rm par}$: $\xi = 0.45-0.5$. In
Figure~\ref{fig:fig11} we show the \agnsph\ case. The points of the other two
simulated sets are similarly located.
For our samples, objects with $M_{par} <-1$ tend to be dynamically
relaxed ($\Gamma <2$) with only few exceptions ($\leq 2$ objects).  
On the other hand, selecting objects with lower values
of $\Gamma$ does not guarantee the X-ray regularity, on the contrary
some clusters with $\Gamma < 2$ have $M_{\rm par}>  3$.

 Other criteria to evaluate the dynamical state have been introduced in the literature such as the center of mass displacement (defined as the offset between the center of mass and the minimum of the potential); the virial ratio between the thermal energy plus the surface pressure term and the kinetic energy; and the substructure mass fraction within the virial radius \citep{neto.etal.07,  power.etal.12,meneghetti.etal.14}.
We checked the performance of $M_{par}$ against the offset parameter derived by \cite{killedar.etal.12} for our SPH set. We verified that also in this case $M_{par}$ is a stronger constraint. We reserve for future investigation a detailed comparison between several dynamical state parameters and the morphological parameters.

The relaxed sample of SPH simulations include 12 clusters that have $M_{par} <-1$ in all of the
three physics. This subsample covers a wide range in mass with
$M_{200}$ spanning from $8 \times 10^{13}$\hinv $M_{\odot}$ to $1.5
\times 10^{13}$\hinv $M_{\odot}$ and it presents the same number of
objects below and above the mass $M_{200}= 4 \times10^{14}$\hinv
$M_{\odot}$.


\bibliographystyle{apj}
\bibliography{ref} 


%

\end{document}